\documentclass[a4paper,fleqn,usenatbib]{mnras}

\usepackage{newtxtext,newtxmath}

\usepackage[T1]{fontenc}
\usepackage{ae,aecompl}

\usepackage{graphicx}	
\usepackage{amsmath}	
\usepackage{amssymb}	
\usepackage{cases}
\usepackage{tabularx}
\usepackage{float}

\title[Perturbations by unseen cold giants]{On the origin of the eccentricity dichotomy displayed by compact super-Earths: dynamical heating by cold giants}

\author[Poon and Nelson]{
Sanson T. S. Poon$^{1,2}$\thanks{E-mail: s.t.s.poon@qmul.ac.uk} and
Richard P. Nelson$^{1}$
\\
$^{1}$Astronomy Unit, Queen Mary University of London, London E1 4NS, UK\\
$^{2}$Royal Observatory Greenwich, London SE10 9NF, UK\\
}

\date{Accepted 2020 September 3. Received 2020 August 14; in original form 2020 June 1.}

\pubyear{2020}

\begin{document}
\label{firstpage}
\pagerange{\pageref{firstpage}--\pageref{lastpage}}
\maketitle

\begin{abstract}
Approximately half of the planets discovered by NASA's Kepler mission are in systems where just a single planet transits its host star, and the remaining planets are observed to be in multi-planet systems. Recent analyses have reported a dichotomy in the eccentricity distribution displayed by systems where a single planet transits compared with that displayed by the multi-planet systems. Using $N$-body simulations, we examine the hypothesis that this dichotomy has arisen because inner systems of super-Earths are frequently accompanied by outer systems of giant planets that can become dynamically unstable and perturb the inner systems. Our initial conditions are constructed using a subset of the known Kepler five-planet systems as templates for the inner systems, and systems of outer giant planets with masses between those of Neptune and Saturn that are centred on orbital radii $2 \le a_{\rm p} \le 10$~au. The parameters of the outer systems are chosen so that they are always below an assumed radial velocity detection threshold of 3~m~s$^{-1}$. The results show an inverse relation between the mean eccentricities and the multiplicites of the systems. Performing synthetic transit observation of the final systems reveals dichotomies in both the eccentricity and multiplicity distributions that are close to being in agreement with the Kepler data. Hence, understanding the observed orbital and physical properties of the compact systems of super-Earths discovered by Kepler may require holistic modelling that couples the dynamics of both inner and outer systems of planets during and after the epoch of formation.
\end{abstract}

\begin{keywords}
planets and satellites: dynamical evolution and stability -- planets and satellites: formation
\end{keywords}



\section{Introduction}\label{sec:intro}
At the time of writing more than 4000 exoplanets have been discovered and confirmed using various detection methods. More than half of them were discovered by the Kepler transit survey \citep{2010Sci...327..977B,2011ApJ...736...19B,2013ApJS..204...24B, 2014ApJS..210...19B,2015ApJS..217...16R,2015ApJS..217...31M, 2016ApJS..224...12C,2018ApJS..235...38T}. Due to detection biases, most of the Kepler planets have short periods $< 100$~days, and less than one percent have orbital periods longer than one year\footnote{All exoplanetary data used in this paper are from \href{https://exoplanetarchive.ipac.caltech.edu}{NASA Exoplanet Archive} unless stated otherwise.}. 
The majority of the Kepler planets have radii between 1 and 4 $\rm{R_{\oplus}}$, and these super-Earths and sub-Neptunes are often found in compact multi-planet systems.

Various analyses of the Kepler data have been undertaken to obtain insight into the formation and evolution of planetary systems. For example, the distribution of the observed planetary system multiplicities shows a sharp increase for single planet systems compared to two planet systems \citep{2011ApJS..197....8L}, and this apparent \emph{Kepler dichotomy} has been interpreted as either arising because of an intrinsic excess of single planet systems \citep{2012ApJ...758...39J}, or alternatively because of the distribution of mutual inclinations within multi-planet systems \citep{2018ApJ...860..101Z}. Understanding the origin and nature of this dichotomy would clearly shed light on the history of formation and dynamical evolution experienced by the compact Kepler systems. An apparent dichotomy has also been detected in the distributions of the orbital eccentricities associated with either single or multiple planet systems \citep{2016PNAS..11311431X, 2019AJ....157...61V, 2019AJ....157..198M}. The Kepler single-planet systems have a mean eccentricity $\langle e_{1}\rangle \approx0.25$--0.3, whereas the multi-planet systems have $\langle e_{\geq 2}\rangle \approx0.05$, indicating the two populations have experienced different dynamical histories.

Different formation and evolution scenarios have been proposed to explain the Kepler compact multi-planet systems. A study of the architectures of multi-planet systems by \citet{2015ApJ...807...44P} showed that the high multiplicity systems are close to being dynamically unstable, and these authors suggested that the low multiplicity systems may have experienced dynamical instabilities and planet-planet collisions, leading to the low numbers of planets now observed in these systems. Dynamical instabilities during the late stages of formation can lead to the self-excitation of eccentricities and mutual inclinations within compact planetary systems, and numerous studies of this process have been undertaken \citep[e.g.][]{2012ApJ...751..158H,2016ApJ...832...34M,2017AJ....154...27M,2017MNRAS.470.1750I,2020MNRAS.491.5595P}. The results of these studies are generally in agreement with the data on the mutual separations between the planets, but when planetary system masses are adopted that are characteristic of those inferred for the Kepler multi-systems, then the degree of gravitational scattering is insufficient to provide large enough mutual inclinations or eccentricities to explain the multiplicity and eccentricity dichotomies described above \citep[e.g.][]{2020MNRAS.491.5595P}.

This has led to an alternative hypothesis for explaining the eccentricity dichotomy, namely that giant planets in the outer regions of planetary systems perturb the inner systems. Observations show that compact systems of super-Earths can have outer giant companions. One example is the Kepler-68 system, which contains two transiting super-Earths/sub-Neptunes (Kepler-68b and c) with orbital periods less than 10 days, and a third planet (Kepler-68d) that was discovered using the radial velocity (RV) technique orbiting beyond 1 au \citep{2013ApJ...766...40G,2014ApJS..210...20M}. In general, cold gas giant planets are on eccentric orbits, and it is well known that planet-planet scattering within a system of gas giants can excite eccentricities to high values ($e>0.3$) \citep{2008ApJ...686..580C,2008ApJ...686..603J,2014ApJ...786..101P}. If a chain of excited outer giants coexists with an inner compact system, then perturbations can excite the eccentricities and mutual inclinations of the inner planets \citep{2013ApJ...767..129M,2017MNRAS.468..549B,2017MNRAS.467.1531H,2017MNRAS.468.3000M,2017AJ....153...42L,2017AJ....153..210H,2018MNRAS.478..197P,2018AJ....156...92Z,2020AJ....159...38M}.

In this paper we explore this idea using $N$-body simulations, similar to previous work \citep[e.g.][]{2017MNRAS.468.3000M,2017AJ....153..210H}, except we explore the effects of adopting a range of multiplicities for the systems of cold giant planets, and we choose parameters for the cold giants such that they would be undetectable in RV surveys that have a detection limit of $v_{\rm RV} = 3~{\rm m~s^{-1}}$. We construct initial conditions that consist of inner systems of super-Earths, based on known 5-planet systems observed by Kepler, and outer systems of giant planets for which the masses are in the range $15 \le m_{\rm p} \le 100$~M$_{\oplus}$ and the semi-major axes are centred between $2 \le a_{\rm p} \le 10$~au. The results of the simulations are processed through a pipeline that synthetically observes the systems using the transit method and produces statistics on the multiplicity and eccentricity distributions for comparison with the Kepler data. Using these tools, we identify regions of parameter space that produce results that are close to being in agreement with the data.

This paper is structured as follows. In Section~\ref{sec:selection} we discuss the selection of outer planetary systems for this study, and in Section~\ref{sec:setup} we describe the set-up of the initial conditions and the selection of the inner planetary system templates. In Section~\ref{sec:results} we present the results of our simulations, and in Section~\ref{sec:syn_obs} we present the synthetic observations of the simulated systems and examine the distributions of multiplicities and eccentricities that arise. In Section \ref{sec:unmodel} we discuss the influence of physical effects such as a realistic collisions model and tidal interactions that were omitted from our primary suite of simulations, and which could potentially affect the final results when comparing to the observations. Finally, we discuss our results and draw conclusions in Section~\ref{sec:discuss}.
\begin{figure}
\centering
\includegraphics[width=1.0\columnwidth]{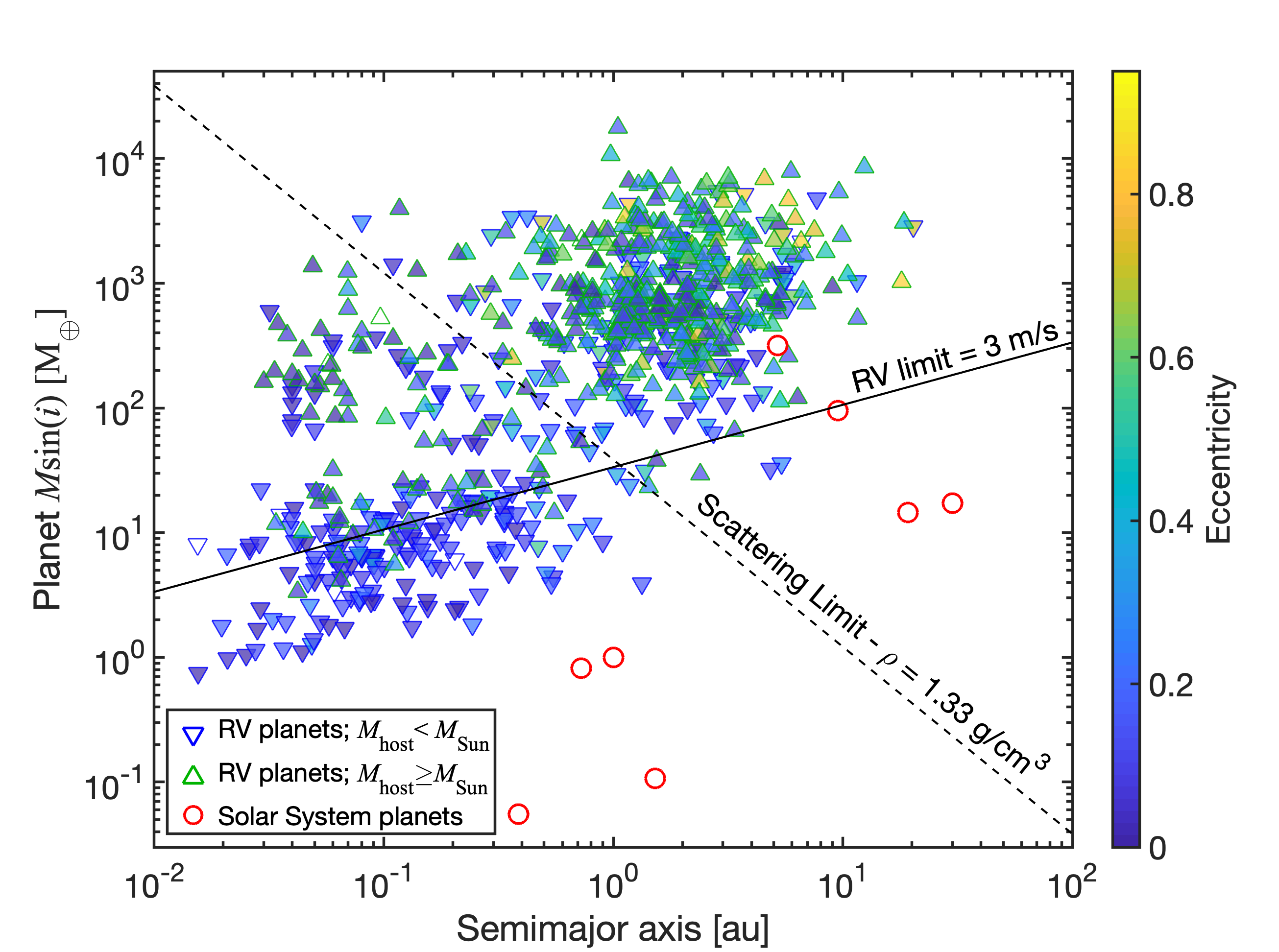}
\caption[RV planets and detection limit]{Masses ($M\sin{i}$) and semi-major axes of confirmed RV planets (triangles), where planets with host star masses $M_{\rm host}< 1$~M$_{\odot}$ are marked with downward pointing triangles, and planets with $M_{\rm host} \ge 1$~M$_{\oplus}$ are marked with upward pointing triangles. The colours of the triangles indicate the eccentricities of the planets (they are left unfilled if the eccentricity is not known). The solid line shows the RV detection limit (equation \ref{eq:RV_limit}) of $v_{\rm RV}=3~{\rm ms^{-1}}$ induced by a planet on an edge-on circular orbit around a Solar type star. The majority of RV planets below the solid line orbit a host star with $M_{\rm host}< 1$~M$_{\odot}$, making detection possible. The dashed line is the scattering limit for a planet given by equation~\ref{eq:scatter_limit}, assuming a Jupiter-like mean density and Solar type host star. Solar System planets are marked using red circles for reference.
}\label{fig:RVlimit}
\end{figure}

\section{Consideration of system selection}\label{sec:selection}
Kepler was inefficient at detecting planets with periods greater than 1 year ($a_{\rm p}\gtrsim1{\rm~au}$) because of detection biases and the mission lifetime. Most of what we know about longer period planets comes from RV surveys. Figure~\ref{fig:RVlimit} shows the masses versus semi-major axes for planets discovered by the RV method, demonstrating the existence of a population of Jovian planets with semi-major axes greater than 1~au. The RV survey presented by \citet{2011arXiv1109.2497M} indicates that approximately 14\% of Solar type stars host a giant planet with orbital period $\lesssim 4000$~days, whereas it has been estimated that up to 50\% of stars host a super-Earth or sub-Neptune with orbital period $\le 100$ days \citep{2013ApJ...766...81F}. Hence it appears that detectable cold giant planets occur less frequently than the warm super-Earths. In this work we are working with the implicit hypothesis that each system of inner planets may be accompanied by a system of outer planets, and hence we need to consider what that unseen population might look like.

Previous studies of inner systems being perturbed by outer planets focused on the RV-like planets, for which cold Jupiters are massive enough to generate a detectable RV signal, $v_{\rm RV}$. The value of $v_{\rm RV}$ is given by
\begin{equation}\label{eq:RV_limit}
v_{\rm RV}=\sqrt{\frac{G}{1-e^{2}}}  \left ( m_{\rm p}\sin{i_{\rm p}} \right ) \frac{1}  {\sqrt{a_{\rm p}(m_{\rm p}+M_{\star})}},
\end{equation} 
where $G$ is the gravitational constant, $m_{\rm p}$ is the mass of the RV planet, $i_{\rm p}$ is the inclination of the planet's orbital plane, $a_{\rm p}$ is the semi-major axis, and $M_{\star}$ is the mass of the host star. We adopt an RV detection limit of $v_{\rm RV}=3~{\rm m~s^{-1}}$, and this limit is indicated in figure~\ref{fig:RVlimit} for an edge on circular orbit ($i_{\rm p}=90^{\circ}$; $e=0$) around a Solar type star ($M_{\star}=1~{\rm M_{\odot}}$). We see that only Jupiter in the Solar system lies above the detection 
limit
while Saturn is just below the limit. 

For outer system planets to be able to induce perturbations on the orbits of the inner system planets, we need to be in the regime where strong planet-planet scattering is favoured over planet-planet collisions. Under the conditions of strong scattering, the velocity kick experienced by a planet relative to a circular Keplerian orbit should correspond approximately to the escape velocity, $v_{\rm e}$, from the perturbing body \citep[e.g.,][]{2008ApJ...686..621F}. Strong scattering can arise when $v_{\rm e}$ is larger than the Keplerian velocity, $v_{\rm K}$ (i.e. $v_{\rm e}/v_{\rm K}>1$). The mass of the planet corresponding to the ratio $v_{\rm e}/v_{\rm K}$ is
\begin{equation}\label{eq:scatter_limit}
m_{\rm p}=\left ( \frac{3}{32\pi} \frac{M^{3}_{\star}}{a^{3}_{\rm p}\rho_{\rm p}}  \right )^{\frac{1}{2}} \left ( \frac{v_{\rm e}}{v_{\rm K}} \right )^{3},
\end{equation} 
where $\rho_{\rm p}$ is the internal mean density of the planet. The dashed line plotted in figure~\ref{fig:RVlimit} shows the scattering limit for planets orbiting a Solar mass star, assuming $\rho_{\rm p}=\rho_{\rm Jupiter}\approx1.33$~${\rm g/cm^3}$, where planets above the line are more likely to induce scattering while the planets below the line are more likely to collide during orbit crossing.

Combining our requirement that the perturbing planets would be undetectable with long term RV surveys, with the need for them to be in the scattering rather than collision regime, means that we need to select bodies that lie in the region bounded by the solid and dashed lines exterior to 1~au in figure~\ref{fig:RVlimit}.

\section{Simulation set-up}\label{sec:setup}
We use the $N$-body code \textsc{mercury} \citep{1999MNRAS.304..793C} to undertake the simulations presented in this paper. Collisions during the simulations are treated using a simple hit-and-stick approach that conserves the total mass and linear momentum when two bodies collide and accrete to form a single object. We have, however, re-run a small subset of simulations to investigate the effects of adopting a more realistic collision model, and tidal interactions with the central star, as these effects were not included in our main suite of simulations. For the initial conditions, we generate templates for the inner and outer planetary systems, and combine them to generate each model, as described below.

\subsection{Inner planetary system templates}\label{subsec:inner}
We have selected a number of Kepler five-planet systems as templates for the inner systems using the following criteria. Similar to \citet{2020MNRAS.491.5595P}, we have selected those Kepler systems where all five of the known planets are transiting. The Kepler-82 and Kepler-122 systems are excluded because Kepler-82f \citep{2019A&A...628A.108F} and Kepler-122f \citep{2014ApJ...787...80H} were discovered by transit timing variations (TTVs). 

We are interested in compact systems of super-Earths, so we have chosen systems in which the known outermost planet has semi-major axis $\le 1.0$ au or orbital period $\le 1$ yr. The Kepler-150 system is excluded as the outermost planet, Kepler-150f, has an orbital period of $\sim 1.74$ yr \citep{2017AJ....153..180S}. Kepler-444 system is also excluded because the planets are too small to be considered as super-Earths \citep{2015ApJ...799..170C}.

\begin{figure}
\centering
\includegraphics[width=1.0\columnwidth]{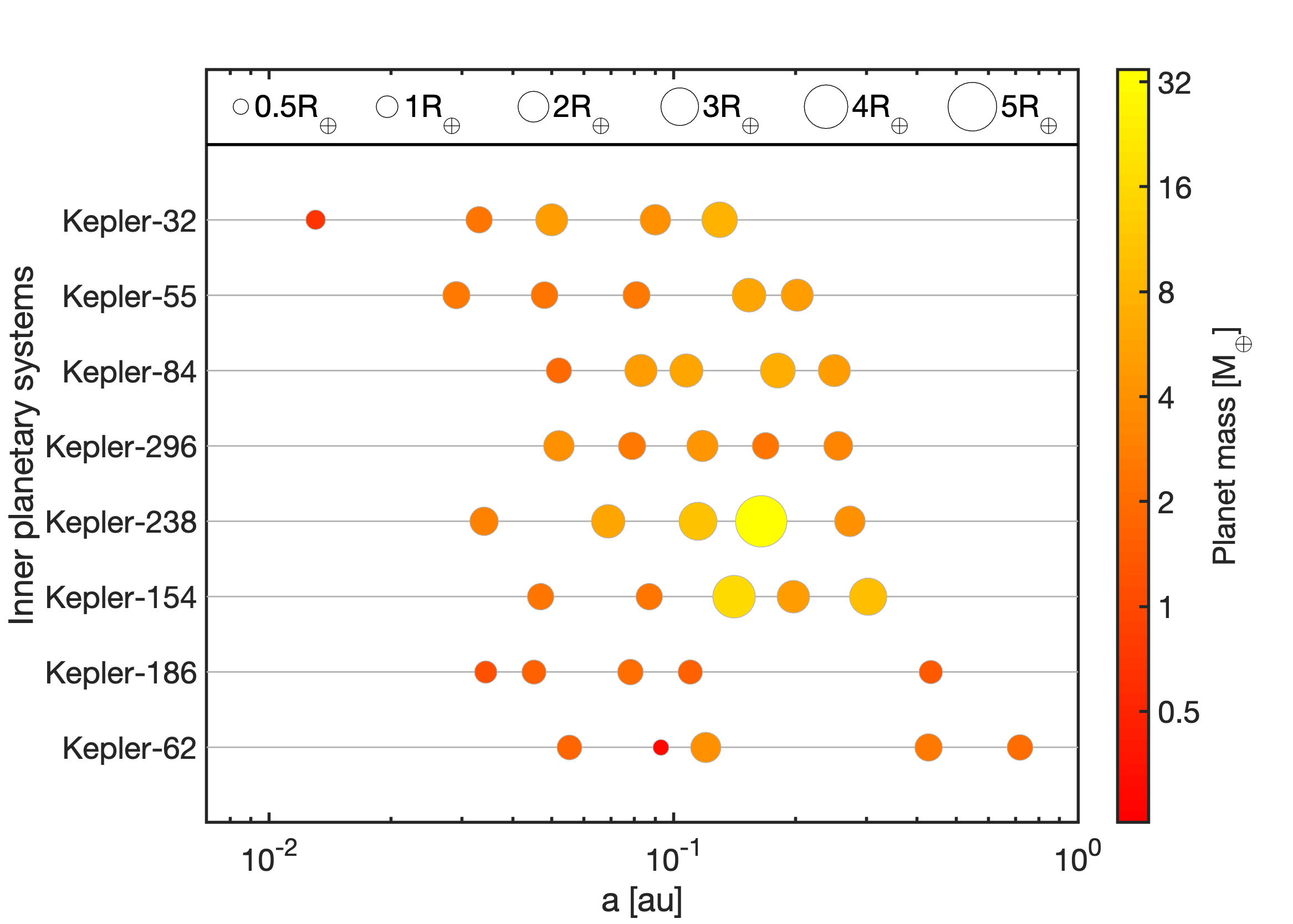}
\caption[Inner planetary templates]{The eight selected inner planetary system templates.  The symbol size represents the radius of each planet and the colour represents the planet mass calculated using the mass-radius relation $M_{\rm p}=R_{\rm p}^{2.06}$ adopted from \cite{2011ApJS..197....8L}.}\label{fig:inner_demo}
\end{figure}

We adopt the mass-radius ($M_{\rm p}-R_{\rm p}$) relation $M_{\rm p}=R_{\rm p}^{2.06}$ suggested by \citet{2011ApJS..197....8L} for the inner planets, where $M_{\rm p}$ and $R_{\rm p}$ are in Earth units. Combining this mass-radius relation with the initial values of the eccentricities and inclinations we adopt (see section \ref{subsec:initialcond}), leads to some of the Kepler systems becoming unstable, with at least one planet experiencing a collision during the simulation. Section~\ref{subsec:stable} describes the results of the stability tests. After excluding the unstable Kepler systems, eight were selected as the inner planetary system templates. They are Kepler-32, -55, -62, -84, -154, -186, -238, and -296 (see figure \ref{fig:inner_demo}).

\subsection{Outer planetary system templates}\label{subsec:outer}
The masses of the outer planets we adopted lie between 15 ${\rm M_{\oplus}}$ and 100 ${\rm M_{\oplus}}$. In contrast to previous studies of the effects of outer giant planets on inner systems \citep[e.g.][]{2015ApJ...808...14M,2017AJ....153..210H,2017MNRAS.468.3000M}, our outer giant planets are below the RV detection threshold discussed above.

The number of outer planets, $N_{\rm out}$, in each simulation is one of 3, 6, or 12. Four different sets of mass are adopted, where $M_{\rm p,out} = 15$ , 30, 60 or 100~${\rm M_{\oplus}}$. In a given simulation, all the outer planets start with the same mass. The mutual separation between each giant planet pair, measured in units of the mutual Hill radius, is $K=4$. The mutual Hill radius for a pair of adjacent planets is defined by
\begin{equation}\label{eq:R_Hi}
R_{{\rm H},i}=\frac{a_{i}+a_{i+1}}{2}\left ( \frac{M_{{\rm p},i}+M_{{\rm p},i+1}}{3M_{\star}} \right )^{\frac{1}{3}},
\end{equation} 
where $a_{i}$ is the semi-major axis of the $i^{\rm th}$ planet in the system. The dimensionless number $K$ is
\begin{equation}\label{eq:Ki}
K=\frac{a_{i+1}-a_{i}}{R_{{\rm H},i}}.
\end{equation}
The value of $K=4$ was chosen as a compromise between ensuring an instability actually occurs during the simulations, having a relatively short instability timescale to make the simulations tractable, and also wanting the instability timescale to not be so short that the outer planetary systems disintegrate within the first few orbits of the simulations. We note a similar value has been used in previous studies \citep{2008ApJ...686..580C,2014ApJ...786..101P,2017MNRAS.468.3000M}. 

Three different sets of locations for the outer planets are considered, with the median semi-major axis for the outer planets, $\tilde{a}_{\rm out}$, being 2, 5, or 10~au. The physical radii of the outer planets, important for determining when collisions occur, are calculated by defining their mean internal densities to be equal to that of Jupiter ($1.33{\rm g/cm^3}$).

\begin{figure}
\centering
\includegraphics[width=1.0\columnwidth]{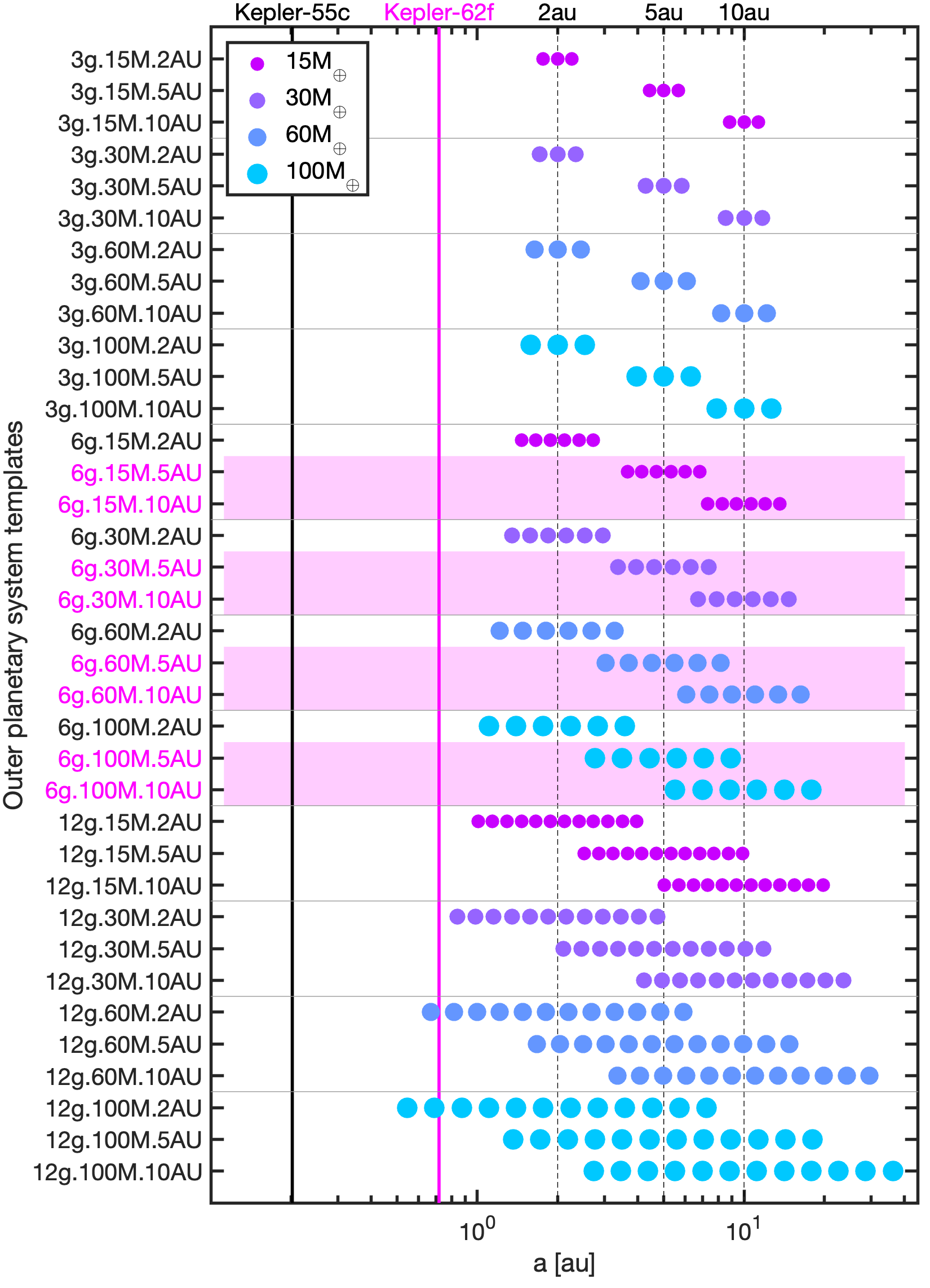}
\caption[Outer planetary templates]{Semi-major axes of the outer planets within different templates (as listed in table~\ref{tab:outertemplate}). Different planet masses are marked in different sizes and colours. The eight outer templates that are the focus of most of our runs (see section~\ref{subsec:focused}) are highlighted in pink. The black solid vertical line marks the semi-major axis of Kepler-55c, which is the outermost planet of the Kepler-55 system. The pink solid vertical line marks the semi-major axis of Kepler-62f, which is the outermost planet within our eight selected inner system templates. Kepler-55 is the system we selected to study with all outer planetary templates, while Kepler-62 (and other selected inner planetary templates) is only combined with the eight outer planetary templates described in the text. There are no immediate dynamical interaction between our inner and outer planetary systems.}\label{fig:outer_template_demo}
\end{figure}

In summary, the parameters for the outer planetary system templates are comprised of three different multiplicities, four different masses and three different values of the median semi-major axis, giving a total of 36 templates that could be generated for each inner planetary system template. We use a labelling convention based on the parameters of the outer planet system when describing the runs as follows: `($N_{\rm out}$)g.($M_{\rm p,out}$)M.($\tilde{a}_{\rm out}$)AU'. For example, \texttt{6g.30M.10AU} refers to the outer planet template with six 30 ${\rm M_{\oplus}}$ planets and a median semi-major axis of 10~au. Table \ref{tab:outertemplate} lists all the outer planetary templates we have considered in this study. Figure~\ref{fig:outer_template_demo} displays the semi-major axis distribution of the outer planetary templates. 
\begin{table}
	\centering
	\caption{The 36 different outer planet templates considered in the study. The name of each template follows the convention `($N_{\rm out}$)g.($M_{\rm p,out}$)M.($\tilde{a}_{\rm out}$)AU'. The eight templates that are the focus of most of our runs (see section \ref{subsec:focused}) are indicated by the star superscript.}
	\label{tab:outertemplate}
	\begin{tabular}{lll} 
		\hline
		\hline
		3 giants  & 6 giants  & 12 giants\\ 
		\hline
		\texttt{3g.15M.2AU} & \texttt{6g.15M.2AU} & \texttt{12g.15M.2AU}\\
		\texttt{3g.15M.5AU} & {\texttt{6g.15M.5AU}}$^{\star}$ & \texttt{12g.15M.5AU}\\
		\texttt{3g.15M.10AU} & {\texttt{6g.15M.10AU}}$^{\star}$ & \texttt{12g.15M.10AU}\\
		\texttt{3g.30M.2AU} & \texttt{6g.30M.2AU} & \texttt{12g.30M.2AU}\\
		\texttt{3g.30M.5AU} & {\texttt{6g.30M.5AU}}$^{\star}$ & \texttt{12g.30M.5AU}\\
		\texttt{3g.30M.10AU} & {\texttt{6g.30M.10AU}}$^{\star}$ & \texttt{12g.30M.10AU}\\
		\texttt{3g.60M.2AU} & \texttt{6g.60M.2AU} & \texttt{12g.60M.2AU}\\
		\texttt{3g.60M.5AU} & {\texttt{6g.60M.5AU}}$^{\star}$ & \texttt{12g.60M.5AU}\\
		\texttt{3g.60M.10AU} & {\texttt{6g.60M.10AU}}$^{\star}$ & \texttt{12g.60M.10AU}\\
		\texttt{3g.100M.2AU} & \texttt{6g.100M.2AU} & \texttt{12g.100M.2AU}\\
		\texttt{3g.100M.5AU} & {\texttt{6g.100M.5AU}}$^{\star}$ & \texttt{12g.100M.5AU}\\
		\texttt{3g.100M.10AU} & {\texttt{6g.100M.10AU}}$^{\star}$ & \texttt{12g.100M.10AU}\\		
		\hline
		\hline
	\end{tabular}
\end{table}

\subsection{Constructing initial conditions for the simulations}\label{subsec:initialcond}
To investigate the effects of the outer ice/gas giant planets on the inner super-Earths, we generate different initial conditions by combining one inner planetary system template and one outer planetary system template. We label this system as `inner-planetary-template.outer-planetary-template'. For example, \texttt{Kepler55.6g.30M.10AU} refers to a run which uses Kepler-55 as the template for the inner system and \texttt{6g.30M.10AU} for the outer system. To provide a statistically meaningful sample, our aim is to run each of these systems 100 times with different random initialisations of the orbital elements. This would involve running 100 simulation for all the combined templates (8 Kepler 5-planet systems $\times$ 36 outer ice/gas giant templates $\times$ 100 runs = 28,800 $N$-body simulations in total), which is not possible given our available computational resources. To make the problem tractable, after undertaking a survey of how all the outer system templates affect one of the inner system templates (Kepler-55), we have selected eight outer planetary templates to focus on in detail as described in section~\ref{subsec:subsecselection}.

Each template is run with 100 different instances of the initial conditions. The initial eccentricities, $e$, of the inner super-Earths are randomly drawn from a Rayleigh distribution with eccentricity parameter, $\sigma_{\rm e}=0.035$. The value of 0.035 is taken from the analysis of the Kepler systems presented by \citet{2019AJ....157..198M}. The initial inclinations, $I$, for each run are randomly drawn from a Rayleigh distribution with inclination parameter, $\sigma_{\rm I}=0.5\sigma_{\rm e}=0.0175$~radians. For the outer planets, the values of initial $e$ and $I$ are uniformly distributed within a range of $0\le e\le 0.07$ and $0\le I\le 0.035$~radians. The distributions follow the relation of $e=2I$, but the initial values of $e$ and $I$ for each planet are independent. The arguments of pericenter, $\omega$, longitudes of ascending node, $\Omega$, and mean anomalies, $M$ are distributed uniformly in the range $0 \leq (\omega,\Omega,M)< 2\pi$.

Objects whose orbital distance exceeds 100~au are removed from the simulations. The central bodies of each system have their masses and radii taken from the Kepler data stored in NASA's exoplanet archive. The time-steps used in the simulations are set to be 1/20th of the shortest orbital period of the system. Each simulation is run for $10^{7}$ yr.

\section{Results}\label{sec:results}

\subsection{Stability of the inner planetary system templates}\label{subsec:stable}
Before embarking on a study of how the inner planetary systems are perturbed by the outer planets, we begin by considering the dynamical stability of the inner system templates in the absence of the outer systems. This provides a control set for the simulations, and demonstrates that the instabilities discussed later in this paper are caused by the outer ice/gas giants. We carried out the stability check for all the selected inner planetary system templates described in section~\ref{subsec:inner}. Adopting the inner eccentricity and inclination distributions mentioned in section~\ref{subsec:initialcond} (i.e. $\sigma_{\rm e}=0.035$ and $\sigma_{\rm I}=0.0175$), the simulations show a number of the Kepler system templates are prone to instability within $10^7$ yr. 

\begin{figure}
\centering
\includegraphics[width=1.0\columnwidth]{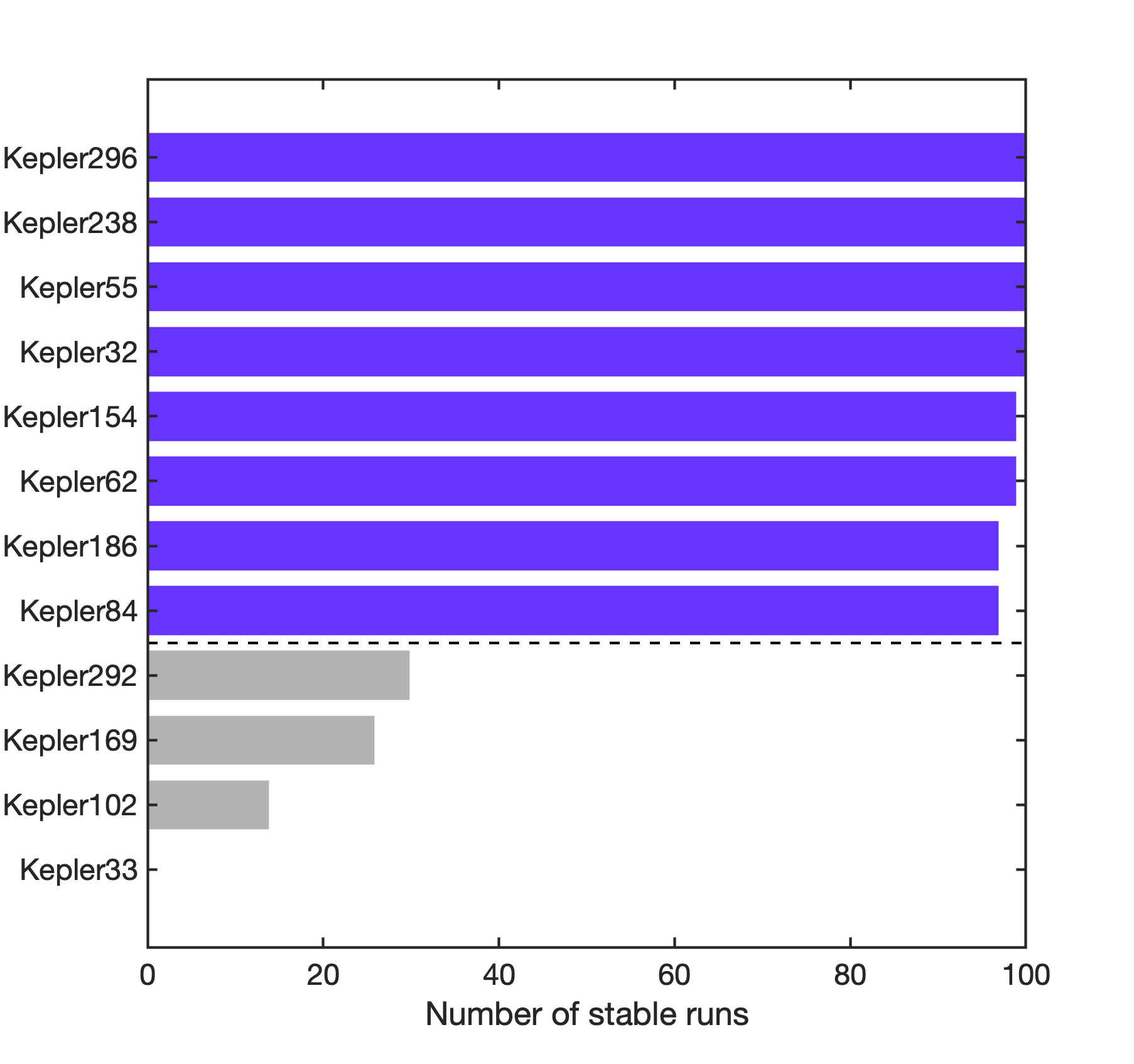}
\caption[Number of stable runs]{Stability control test for the selected Kepler 5-planet inner system templates. The histogram shows the number of stable runs for each template out a total of 100 simulations. Systems with high levels of stability within $10^7{\rm~yr}$ are marked by a purple bar, and the rest are marked in grey. }
\label{fig:stablenumber}
\end{figure}

Figure \ref{fig:stablenumber} shows the results for the stability test. 100\% of the simulations for \texttt{Kepler32}, \texttt{55}, \texttt{238}, and \texttt{296} were stable, while $\geq 97\%$ of the simulations for \texttt{Kepler62}, \texttt{84}, \texttt{154}, and \texttt{186} were stable. The simulations for \texttt{Kepler33}, \texttt{102}, \texttt{169}, and \texttt{292} showed higher levels of instability, and for this reason we remove these systems from further consideration. We note that the purpose of this study is not to specifically assess the stability of the Kepler 5-planet systems, since we have made assumptions about the mass-radius relationship and the initial eccentricities and inclinations that may not apply to each of the systems separately. Instead, our aim is to obtain a sample of stable inner systems, given our assumptions, that can then be evolved in the presence of outer systems of giants to examine whether or not the induced perturbations lead to inner systems similar to those that have been observed.

From now on we consider only the eight inner systems: \texttt{Kepler32}, \texttt{55}, \texttt{62}, \texttt{84}, \texttt{154}, \texttt{186}, \texttt{238} and \texttt{296}. The K-values, that quantify the mutual separations between planets in these systems, are listed in table~\ref{tab:Kval}, along with the stellar masses.
The minimum value in the table is $K_{\rm min}=10.3$, which is greater than the $K_{\rm min}= 7.1$ required to ensure stability in a 5-planet system with circular orbits for up to $10^6$ years \citep{2019MNRAS.484.1538W}, in agreement with the results of our stability tests.

\begin{table}
	\centering
	\caption{$K$-values of the stable Kepler 5-planet system templates. $K_1$ to $K_4$ denote the $K$-value from the 1st to 4th pair of adjacent planets, respectively, in order of increasing orbital radius. $\langle K \rangle$ denotes the arithmetic mean of $K$ for the system. Starred values are the minimum $K$-values in the system, $K_{\rm min}$. The superscripts refer to the following references for the stellar masses: a) \citet{2013MNRAS.428.1077S}; b) \citet{2013ApJS..204...24B}; c) \citet{2014ApJS..210...25X}.}
	\label{tab:Kval}
	\begin{tabular}{lcccccl} 
		\hline
		\hline
		System & $K_{1}$ & $K_{2}$ & $K_{3}$ & $K_{4}$ & $\langle K \rangle$ & $M_{\odot}$\\ 
		\hline
		Kepler-32  & 50.4 & 17.5 & 22.7 & {13.3}$^{\star}$ & 26.0 & 0.58\\
		Kepler-55  & 23.9 & 25.6 & 26.7 & {10.3}$^{\star}$ & 21.6 & 0.62$^{\rm a}$\\
		Kepler-62  & 35.4 & {13.9}$^{\star}$ & 52.8 & 26.8 & 32.2 & 0.69\\
		Kepler-84  & 24.6 & {11.6}$^{\star}$ & 21.0 & 13.9 & 17.8 & 1.00\\
		Kepler-154 & 34.5 & 17.4 & {10.9}$^{\star}$ & 16.4 & 19.8 & 0.89$^{\rm b}$\\
		Kepler-186 & {15.8}$^{\star}$ & 28.5 & 18.0 & 66.9 & 32.3 & 0.54\\
		Kepler-238 & 33.7 & 20.3 & {10.8}$^{\star}$ & 14.9 & 19.9 & 1.06$^{\rm c}$\\
		Kepler-296 & 17.2 & 16.2 & {14.8}$^{\star}$ & 18.0 & 16.5 & 0.50\\
		\hline
		\hline
	\end{tabular}
\end{table}

\subsection{Perturbations by the outer ice/gas giants}\label{subsec:subsecselection}
It would be time consuming and beyond our computational capabilities to run 100 instances for all combinations of the inner and outer planetary system templates. To optimise the computational resources, we selected one of the eight stable inner templates (\texttt{Kepler55}) and ran simulations combining this with all the outer system templates listed in table \ref{tab:outertemplate}. We ran 100 simulations for each configuration. The purpose here is to determine which of the outer planet system configurations give the most promising results, so that we can then focus on these systems in a more comprehensive study.

\begin{figure}
\centering
\includegraphics[width=1.0\columnwidth]{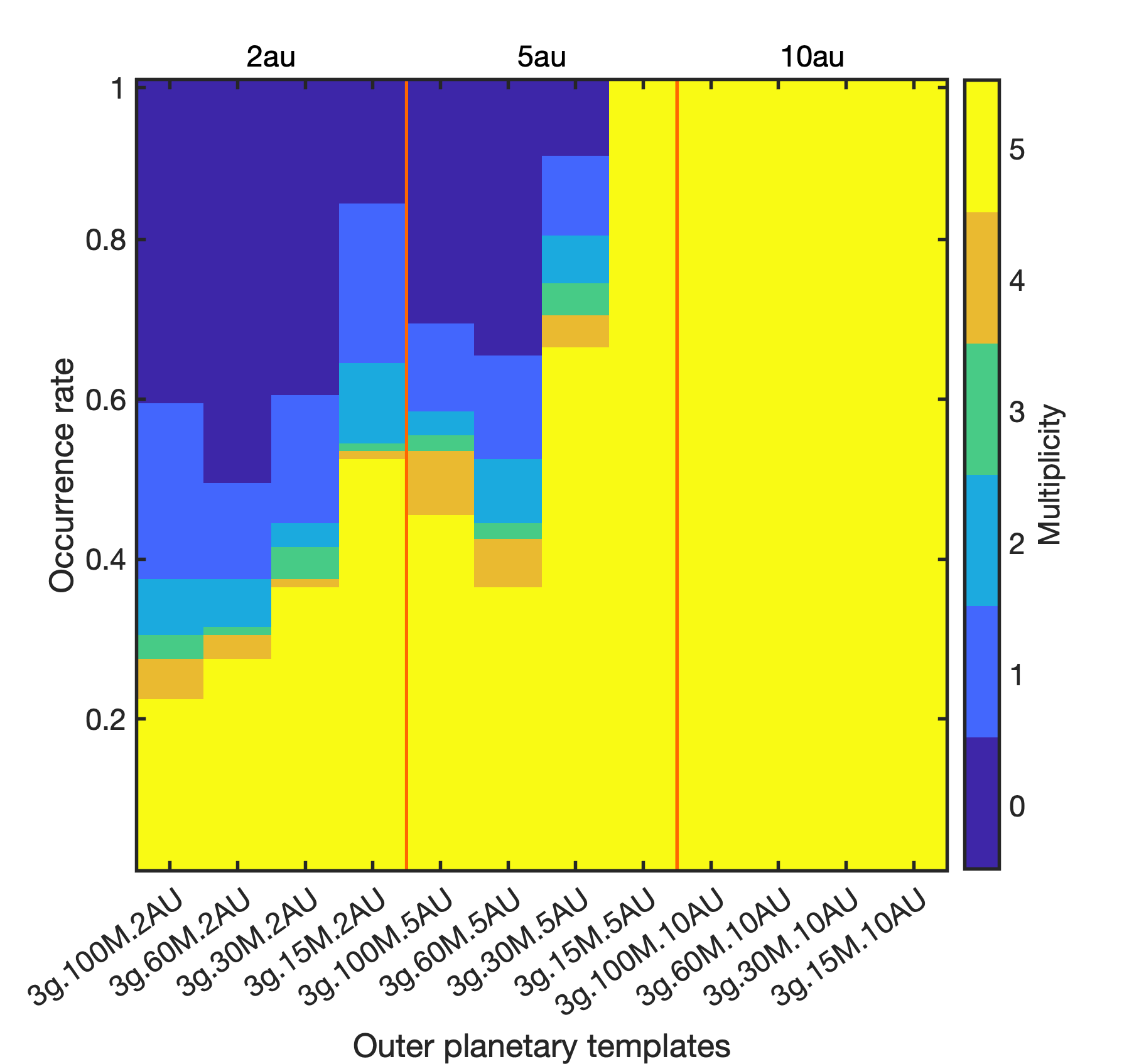}
\caption[Multiplicities for $N_{\rm out}=3$]{Occurrence rate for the final inner planet multiplicities for $N_{\rm out}=3$ (\texttt{3g}) runs on \texttt{Kepler55}. The left panel is the templates with $\tilde{a}_{\rm out}=2 {\rm~au}$, middle panel is the templates with $\tilde{a}_{\rm out}=5 {\rm~au}$, and right panel is for the $\tilde{a}_{\rm out}=10 {\rm~au}$ templates. Different colour represent their relative final multiplicity. Multiplicity of 5 (yellow bar) illustrate unperturbed system and 0 (dark blue bar) for completely destroyed system.}\label{fig:planet_no_3g}
\end{figure}

\begin{figure}
\centering
\includegraphics[width=1.0\columnwidth]{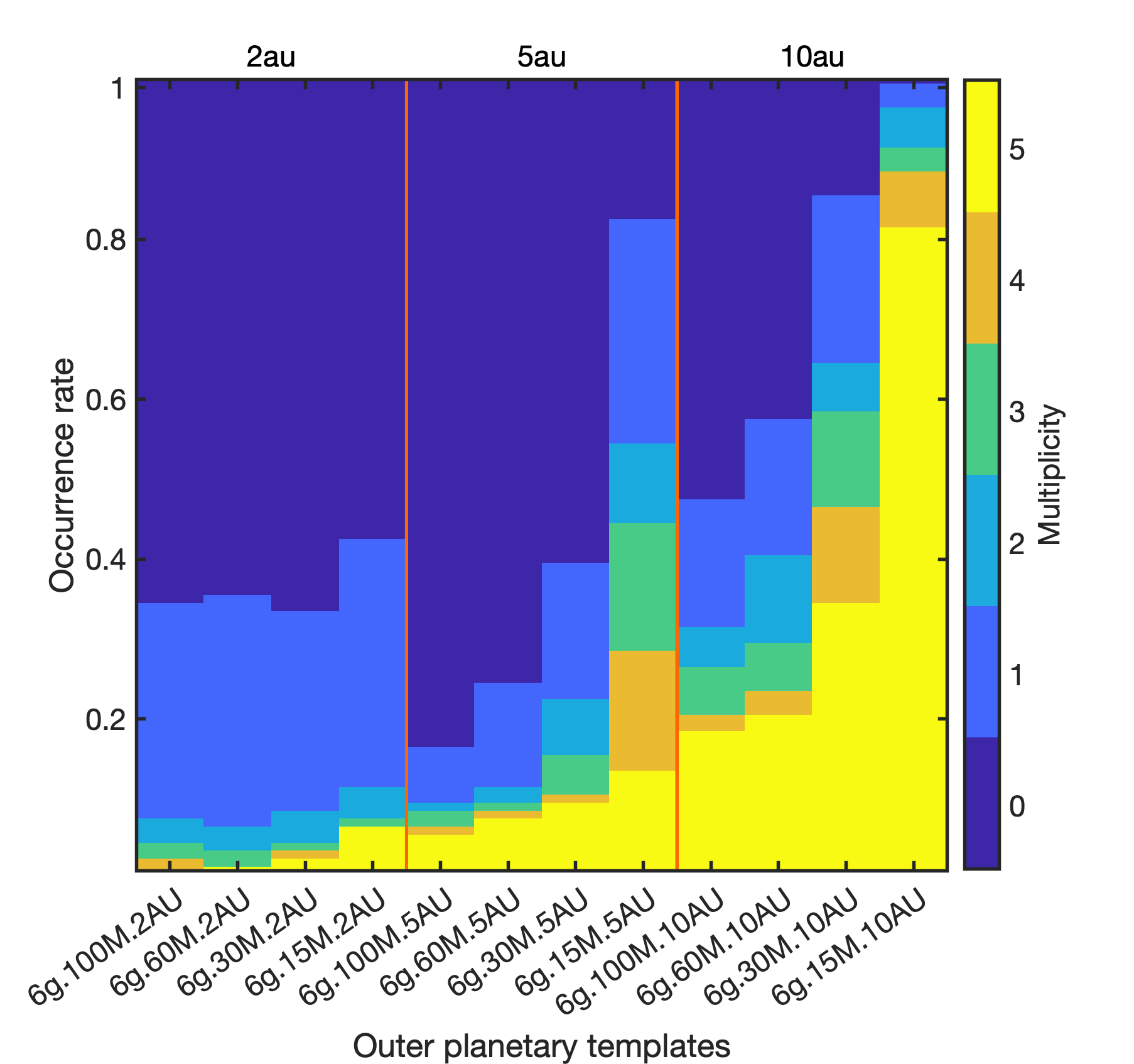}
\caption[Multiplicities for $N_{\rm out}=6$]{Same as figure \ref{fig:planet_no_3g} but for $N_{\rm out}=6$ (\texttt{6g}) outer templates.}\label{fig:planet_no_6g}
\end{figure}

\begin{figure}
\centering
\includegraphics[width=1.0\columnwidth]{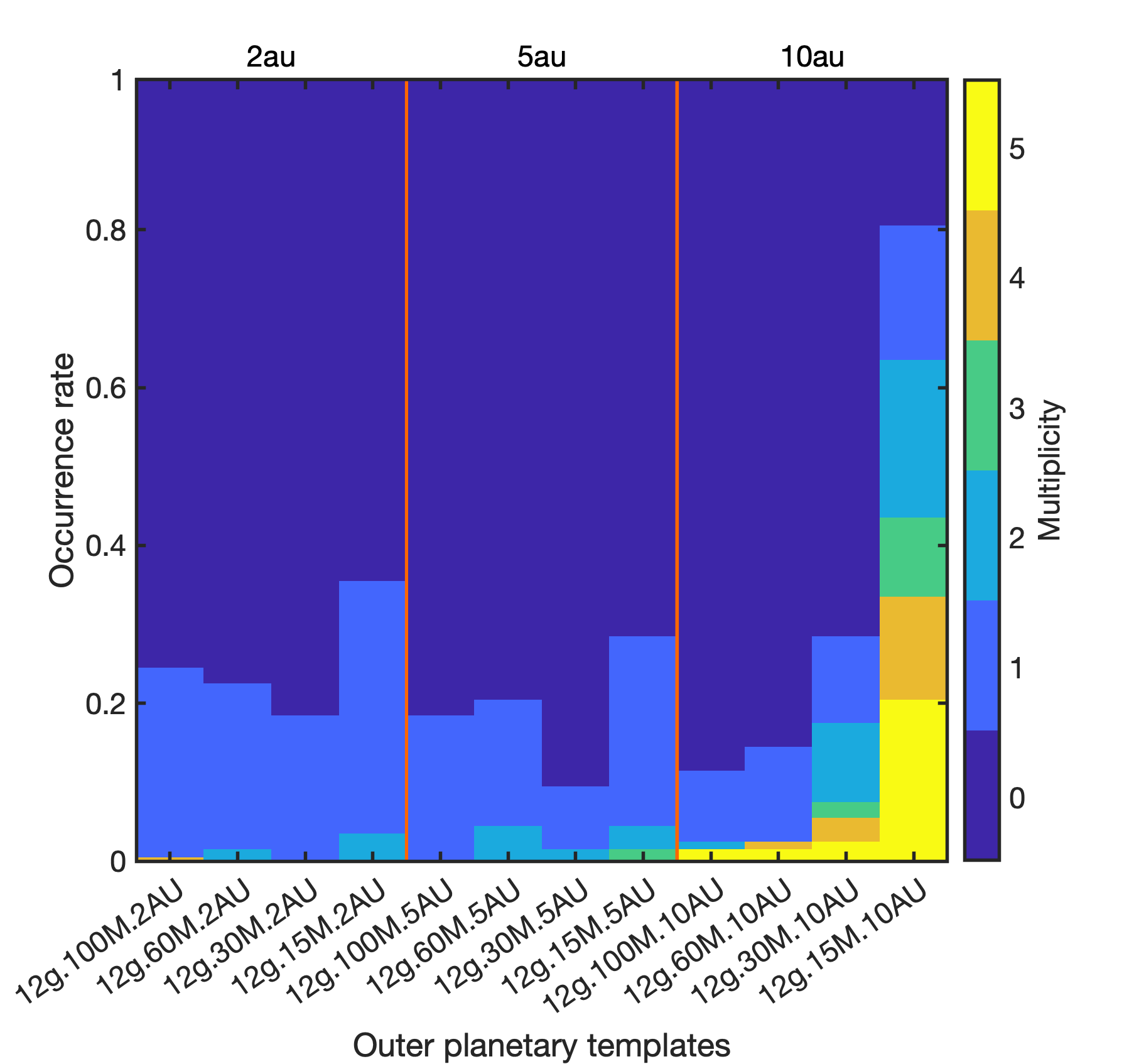}
\caption[Multiplicities for $N_{\rm out}=12$]{Same as figure \ref{fig:planet_no_3g} but for $N_{\rm out}=12$ (\texttt{12g}) outer templates.}\label{fig:planet_no_12g}
\end{figure}

The final inner system multiplicities from this set of simulations are shown in figures~\ref{fig:planet_no_3g}, \ref{fig:planet_no_6g}, and \ref{fig:planet_no_12g}, which show the results from the runs with 3, 6 and 12 outer planets, respectively.
In figure \ref{fig:planet_no_3g} we can see a drop in the final multiplicities as a function of $\tilde{a}_{\rm out}$. Furthermore, for a given value of $\tilde{a}_{\rm out}$, we can also see a decrease in final multiplicities as the masses of the outer planets increases. As expected, larger planet masses induce a greater degree of scattering, and closer orbits increase the probability of outer planets having close encounters with the inner planets systems. The same behaviour is also seen in figures~\ref{fig:planet_no_6g} and \ref{fig:planet_no_12g}.

For the $N_{\rm out}=3$ templates (figure~\ref{fig:planet_no_3g}), the subset with $\tilde{a}_{\rm out}=10$~au shows no perturbation of the final multiplicities of the inner systems at all. Hereafter, we refer to these kinds of systems, where the initial and final values of inner multiplicity are the same ($N_{\rm in,init}=N_{\rm in,final}$), as `unperturbed systems'. The \texttt{3g.15M.5AU} template also produces unperturbed systems only. The remaining outer templates show that the occurrence rate of unperturbed systems is between $\sim$30\% to $\sim$75\%, while the occurrence rate of `completely destroyed systems' ($N_{\rm in,final}=0$) is in the range of $<$10\% to $\sim$50\%.

The $N_{\rm out}=6$ templates (figure~\ref{fig:planet_no_6g}) show a very different occurrence rate of unperturbed and completely destroyed systems compared to the $N_{\rm out}=3$ templates, and more generally there is a very obvious trend towards greater degrees of perturbation of the inner systems as the number of planets in the outer systems increases. For $N_{\rm out}=6$, the occurrence rate of completely destroyed systems covers a wide range between $\sim$20 to 100 per cent, while the unperturbed systems also shows a wide variety, from 0 to $\sim$80 \%. Meanwhile, the proportion of final multiplicities being equal to 2, 3, and 4 are also higher when comparing to the $N_{\rm out}=3$ templates. Figure~\ref{fig:planet_no_12g} shows the final multiplicities for the $N_{\rm out}=12$ templates. Most of these templates resulted in $\sim$80 \% of the systems being completely destroyed. The templates with $\tilde{a}_{\rm out}=10{\rm~au}$ are the only ones that result in unperturbed and moderately perturbed systems among the $N_{\rm out}=12$ templates.

The results shown in figures~\ref{fig:planet_no_3g}, \ref{fig:planet_no_6g}, and \ref{fig:planet_no_12g} agree with our expectations. Much stronger perturbations are experienced by the inner systems when the outer system planets are: i) closer; ii) more massive; iii) more numerous. The probability of having a close encounter between inner and outer planets is obviously larger for closer in outer systems, and greater degrees of scattering are expected when the outer planets are either more numerous or more massive.

Based on the results shown in figures~\ref{fig:planet_no_3g}, \ref{fig:planet_no_6g}, and \ref{fig:planet_no_12g}, we have selected systems with six outer planets and with $\tilde{a}_{\rm out}=5$ and 10~au for a more in-depth study. As highlighted in table~\ref{tab:outertemplate} and figure~\ref{fig:outer_template_demo}, in total this sample contains eight outer system templates: \texttt{6g.15M.5AU}, \texttt{6g.30M.5AU}, \texttt{6g.60M.5AU}, and \texttt{6g.100M.5AU} for $\tilde{a}_{\rm out}=5$~au, and \texttt{6g.15M.10AU}, \texttt{6g.30M.10AU}, \texttt{6g.60M.10AU}, and \texttt{6g.100M.10AU} for $\tilde{a}_{\rm out}=10$~au. The reason for choosing these systems is that they covered the widest range of final multiplicities in our previously described runs (see figure \ref{fig:planet_no_6g}). For example, all the runs ofor \texttt{6g.100M.5AU} show a perturbed inner system, while \texttt{6g.15M.10AU} shows a majority of unperturbed systems. 

\begin{figure*}
\centering
\includegraphics[width=1.0\textwidth]{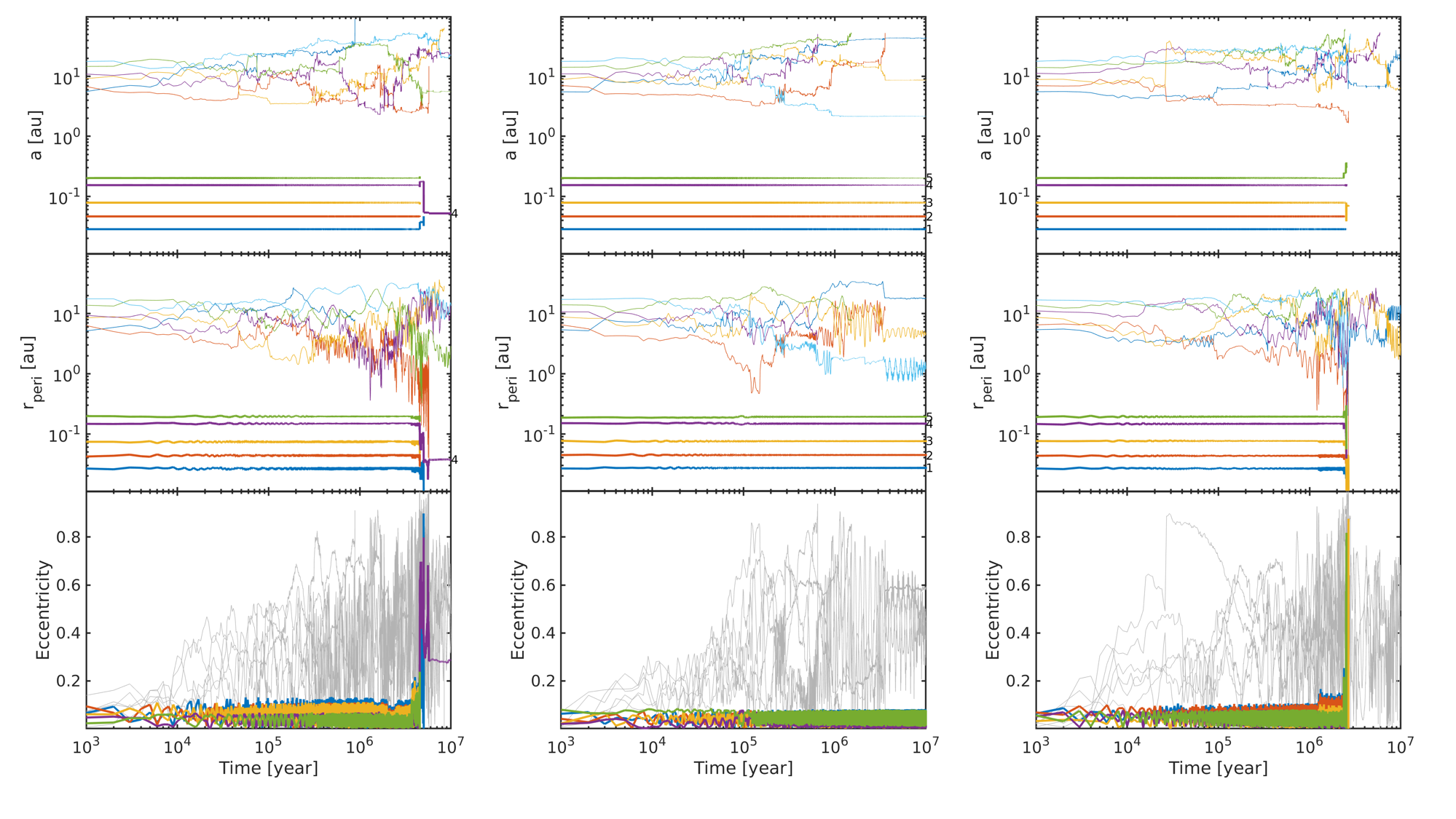}
\caption[Dynamical evolution]{Dynamical evolution during three different runs for the \texttt{Kepler55.6g.60M.10AU} template. Top panel: evolution of planet semi-major axes. Middle panel: evolution of the pericentre distances. Bottom panel: evolution of the eccentricities. Inner system planets are marked in thick solid lines, the numbers labelling the final remaining planets are marked on the right hand side of each plot. Outer system planets are marked using thin solid lines. The left panels show a system in which only one eccentric inner planet survives ($e_{\rm p}\sim0.3$). The middle panels show an unperturbed system in which all five inner planets survive. The right panels show a completely destroyed system.
}\label{fig:arpe}
\end{figure*}

\subsection{Evolution of Kepler templates with 6-planet outer systems}\label{subsec:focused}
As described in section~\ref{subsec:subsecselection}, the templates consisting of 6 outer planets centred around both 5 and 10~au provide the eight templates we investigate further. Together with the eight Kepler inner system templates described in section~\ref{subsec:stable}, there are 64 systems that we now focus on. For each system we run 100 simulations.

\subsubsection{Dynamical evolution}\label{subsubsec:dynamic}
The perturbations arising from the chaotic dynamics of the outer planets can lead to very diverse outcomes for a given combination of inner and outer system templates. Figure~\ref{fig:arpe} shows the evolution for three different runs from the \texttt{Kepler55.6g.60M.10AU} system, where the top panel shows the semi-major axis of the planets, the middle panel shows the distance of the pericentre of the planet's orbit to the host star, $r_{\rm peri}$, and the bottom panel shows the evolution of the eccentricities. 

The left column shows a run where a single planet survives in the inner system. Approximately 1000 years after the start of the simulation, the outer system becomes unstable. This is as expected and compares well to the instability timescales obtained by \citet{2008ApJ...686..580C} when $K=4$. As the eccentricities of some of the outer planets increase, their pericentre distances decrease and start to approach the inner planets. At $t= \times 10^6$ yr one of the outer planets has $r_{\rm peri}<0.4$~au (thin orange line), strongly perturbing the inner system. Two inner planets are scattered to high eccentricities ($e_{\rm p}\gtrsim0.8$), resulting in a collision that forms a single planet with $e_{\rm p} \sim0.3$. The remaining inner planets collide with the giant interloper, an outcome that is common during the simulations. If we were to observe the final state of this system using a transit survey and a viewing position which allows the remaining inner planet to transit its host star, we would classify this system as a being a single, high eccentricity super-Earth. The outer planets would not generally be detectable because of their orbital inclinations with respect to the inner system, and because their orbital periods are too long for multiple transits to be detected within the few years of operation of a Kepler-like survey. Furthermore, the outer planets are below the detection threshold of an RV survey.

The middle column of figure~ \ref{fig:arpe} shows the dynamical evolution of an unperturbed system. Even through we label it as unperturbed, the outer planets still experience chaotic evolution due to the small initial Hill separations. The difference compared to the previously described run is simply that the outer planets in this simulation did not make an excursion into the inner system during the chaotic phase. The value of $r_{\rm peri}$ shows that one of the outer planets got as close as $\sim0.5$ au (thin orange line), while the outermost inner planet is sitting at $\sim0.2$ au (thick green line). The Hill separation of this planet pair was greater than 18 throughout the simulation, leading to only a small perturbation of the inner system, as demonstrated by the evolution of the eccentricities that remain at essentially their initial values. Again, if we were to observe the final state of this system using a transit survey, there would be a finite chance of detecting it as a multi-planet system with low eccentricities, depending on the viewing angle. Together with the high eccentricity single planet system that we discussed above, these two systems demonstrate how different evolutionary paths originating from similar initial conditions can in principle explain results such as those presented by \citet{2016PNAS..11311431X}, for which the mean eccentricity of single-systems is $\langle e_{1} \rangle \approx 0.25$-0.3 and for multi-systems is $\langle e_{\ge2} \rangle \approx 0.05$.

An example of evolution leading to a completely destroyed system is shown in the right column of figure~\ref{fig:arpe}. Similar to the run shown in the left column, the chaotic evolution in the outer system reduces $r_{\rm peri}$ so that an outer planet penetrates into the inner system at $\sim 2.5 \times 10^6$ yr. This induces an instability within the inner system, and allows one of the inner planets (thick yellow line) to accrete the other four inner planets before colliding with the host star.

Overall in our simulations, the dominant mechanism that generates perturbed inner systems is strong scattering of outer giants, leading to one or more giant planet being scattered sufficiently that it passes within the inner system during pericentre passage ($>95 \%$ of the perturbed systems show orbit crossing involving an outer giant and the inner system). Some previous studies have suggested that perturbations of inner system can arise because of secular interactions \citep[e.g. ][]{2013ApJ...767..129M}, but this mechanism is not commonly seen in our study.

\subsubsection{Multiplicities and eccentricities of the inner systems}\label{subsubset:no_and_e_inner}
\begin{figure}
\centering
\includegraphics[width=1.0\columnwidth]{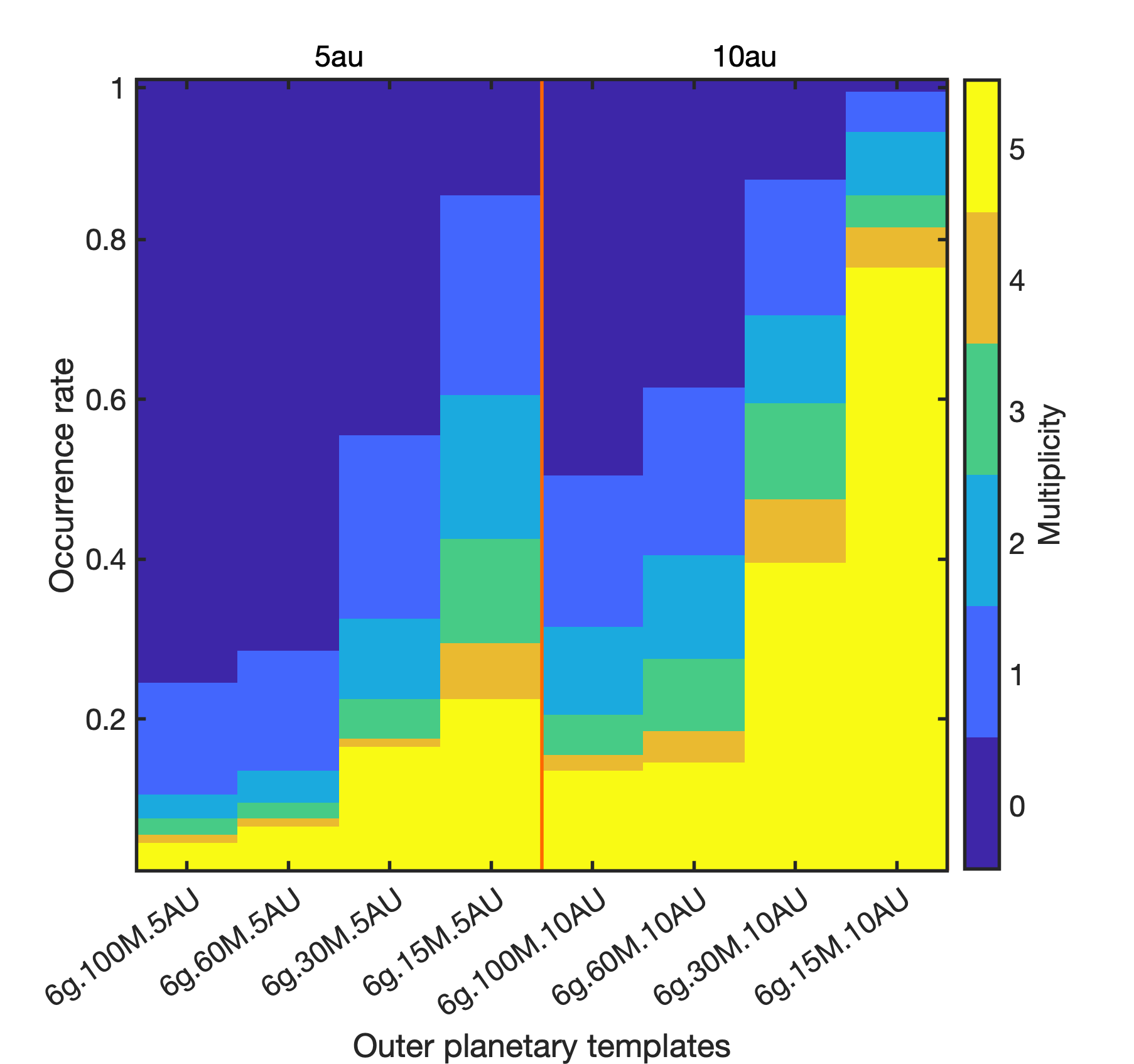}
\caption[Multiplicity occurrence rates of the focused templates]{Multiplicity occurrence rates from the simulations of the 8 selected outer system templates combined with all the Kepler templates. The left-most histograms are for systems with $\tilde{a}_{\rm out}=5 {\rm~au}$, and right-most histograms are for the $\tilde{a}_{\rm out}=10 {\rm~au}$.}\label{fig:planet_no_all_template}
\end{figure}

Before we undertake synthetic observations of the planetary systems resulting from the $N$-body simulations, and compare them with the observations, we discuss some of their intrinsic properties. We find the multiplicities and eccentricities in the final planetary systems to be very diverse. Figure~\ref{fig:planet_no_all_template} shows the final multiplicities obtained from each of the templates, and these can be compared to the 5 and 10~au subsets in figure~\ref{fig:planet_no_6g} which apply only to the \texttt{Kepler55} system. The overall multiplicities obtained from all 8 inner system templates are very similar to those obtained from the \texttt{Kepler55} system, indicating that the dynamics of the outer systems are the main controller of the evolution of the inner systems. 

The trend in the multiplicities displayed by the different outer system templates is not unexpected. Planet-planet scattering is more effective for systems containing more massive planets, and planets in the outer systems are more easily able to penetrate into the inner systems if they orbit closer to the central star. Hence, the correlations shown in figure~\ref{fig:planet_no_all_template} between final multiplicities of the inner systems and the properties of the outer systems are easily understood. The \texttt{6g.15M.10AU} template shows the largest fraction of unperturbed systems at $\sim 75\%$, and the \texttt{6g.100M.5AU} template produced the largest fraction of inner systems that were completely destroyed system, again $\sim 75\%$ of the total. The \texttt{6g.15M.5AU} template produced the largest fraction of systems that were perturbed but not completely destroyed (i.e. $1\le N_{\rm in, final} \le4$). 

\begin{figure*}
\centering
\includegraphics[width=1.0\textwidth]{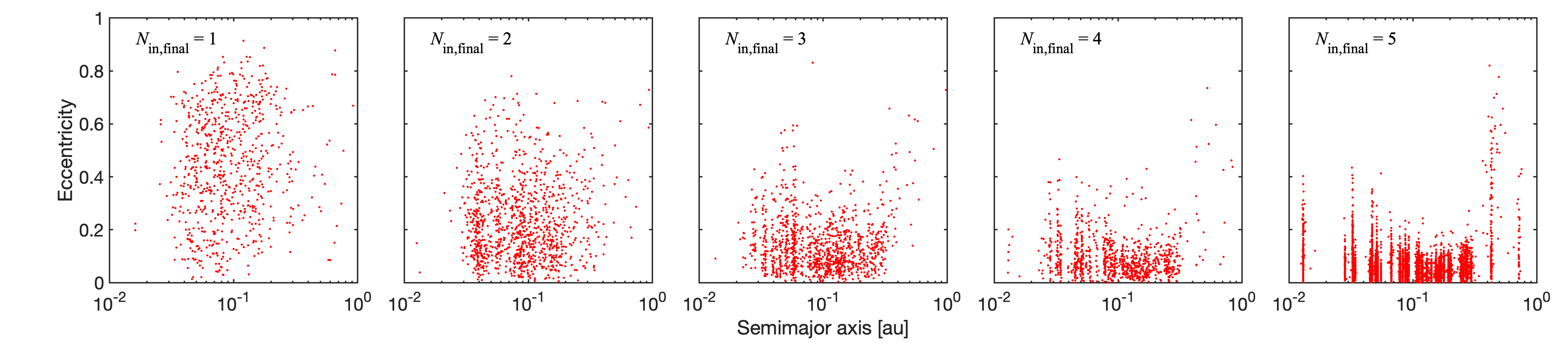}
\caption[Eccentricity-semimajor axis distributions of inner planetary systems]{Scatter plots of $e$ as a function of $a$ for the final inner planetary systems. Different panel shows the distribution of different final inner system multiplicities (from the left panel for $N_{\rm in,final}=1$ in order to right panel for $N_{\rm in,final}=5$).}\label{fig:inner_e_subplot}
\end{figure*}

\begin{figure}
\centering
\includegraphics[width=1.0\columnwidth]{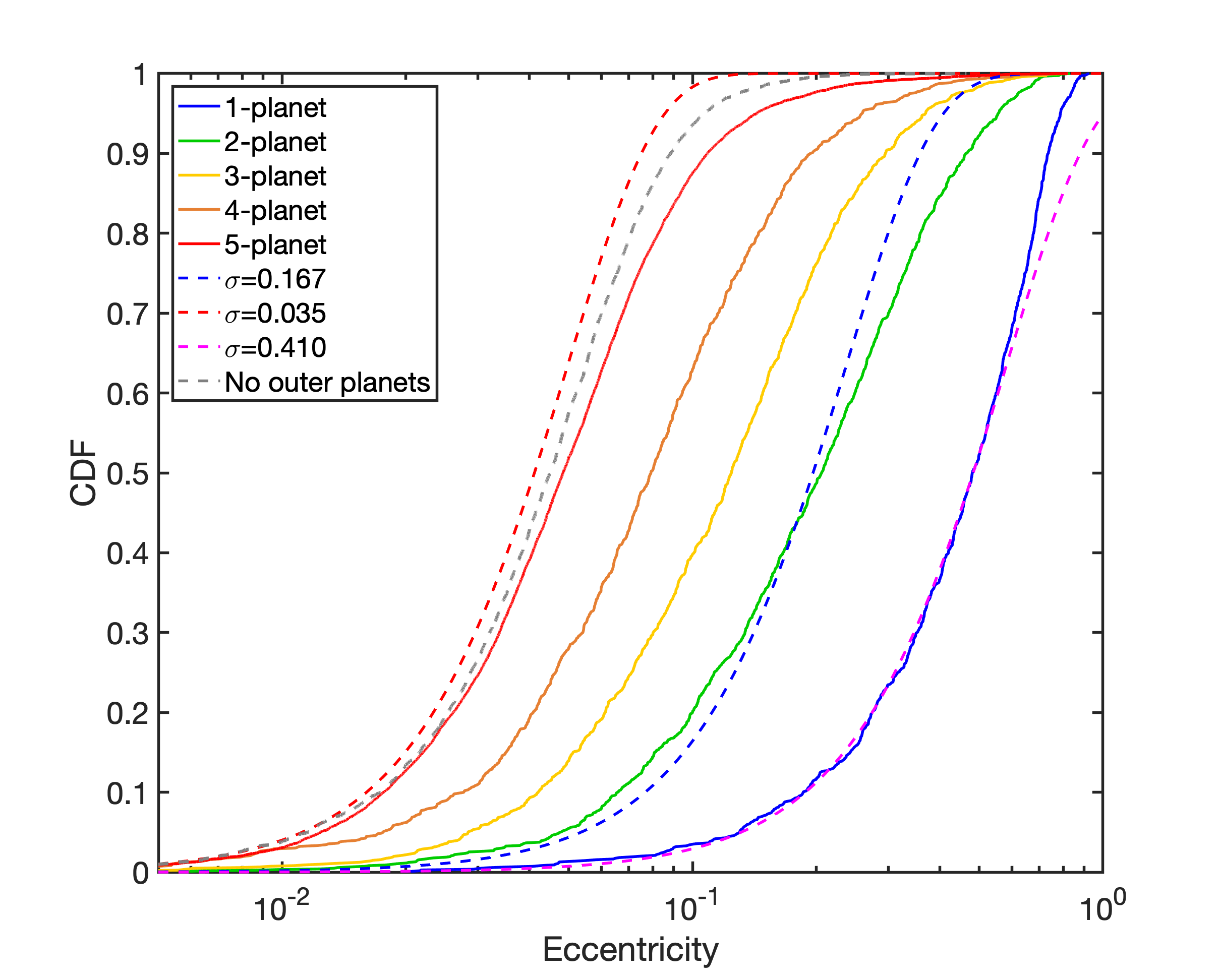}
\caption[Distribution of final eccentricities]{Cumulative distribution functions, CDFs, of the final inner system planet eccentricities obtained from the focused templates simulations. The solid lines correspond to the distribution of different $N_{\rm in,final}$-planet systems. For comparison, the blue, red, and purple dashed line show the distribution of eccentricities drawn from Rayleigh distributions with eccentricity parameters $\sigma = 0.167$, $0.035$, and $0.410$ respectively. The gray dashed line is the CDF drawn from the set of control (with no outer planetary system, section \ref{subsec:stable}).}\label{fig:e_sim}
\end{figure}

The final multiplicities of the inner systems at the ends of the simulations reflect different dynamical histories, and in general one might expect higher multiplicities to arise in systems that have experienced smaller perturbations from the outer planets. One might also therefore expect the final eccentricity distributions to correlate with the multiplicities.
Figure~\ref{fig:inner_e_subplot} shows eccentricities versus semi-major axes from all 64 inner/outer template combinations, with each panel showing data for the different multiplicities (one to five planets). 
Table \ref{tab:e_mean} also lists the mean and median eccentricities, $\langle e \rangle$ and $\tilde{e}$, for different multiplicities at the ends of the simulations. These values, and figures~\ref{fig:inner_e_subplot} and \ref{fig:e_sim}, show a clear inverse relation between the final multiplicities and the values of $\langle e \rangle$ or $\tilde{e}$, as expected.
Figure~\ref{fig:e_sim} shows the final eccentricity distribution for the unperturbed systems (5-planets, red solid line) has remained similar to the initial eccentricity distribution (Rayleigh distribution with $\sigma_{e}=0.035$, red dashed line). This can also be compared to the eccentricity distribution that arose from the simulations performed to check the stability of the Kepler templates, described in section~\ref{subsec:stable} (denoted by the grey dashed line in figure~\ref{fig:e_sim}). The similarity between these three distribution shows that if the inner system retains the original multiplicity, the perturbations from the outer systems are small and do not significantly excite the inner systems. 

\begin{table}
	\centering
	\caption{Mean ($\langle e \rangle$), median ($\tilde{e}$), and standard deviations ($\sigma$) of the eccentricity of planets in the final inner systems from the focused outer templates runs. Different column represent different final inner multiplicities. The eccentricity distributions with values of $\langle e \rangle$, $\tilde{e}$ , and $\sigma$ listed here are show in figure \ref{fig:inner_e_subplot} and \ref{fig:e_sim}.}
	\label{tab:e_mean}
	\begin{tabular}{crrrrr} 
		\hline
		\hline
		& \multicolumn{5}{c}{\underline{Multiplicity}} \\
		& 1 & 2 & 3 & 4 & 5 \\ 
		\hline
		$\langle e \rangle$ 		& 0.48 & 0.24 & 0.15 & 0.10 & 0.06\\
		$\tilde{e}$  	& 0.48 & 0.21 & 0.12 & 0.08 & 0.05\\
		$\sigma$  		& 0.21 & 0.16 & 0.11 & 0.09 & 0.06\\
		\hline
		\hline
	\end{tabular}
\end{table}

The single planet systems (blue solid line in figure~\ref{fig:e_sim} and left panel in figure~\ref{fig:inner_e_subplot}) are the most eccentric on average. The $e$-distribution for 1-planet systems is well fitted by a Rayleigh distribution with $\sigma_{e}=0.410$ for $e<0.6$ (figure~\ref{fig:e_sim}, magenta dashed line), but these systems did not provide enough high eccentricity planets with $0.6<e<1$ to fit the large eccentricity end of the Rayleigh distribution. The $e$-distribution for 2-planet systems (green solid line) is similar to the Rayleigh distribution with $\sigma_{e}=0.167$ (blue dashed line, the suggested Rayleigh distribution eccentricity parameter for observed single-planet system suggested by \citet{2019AJ....157..198M}). We note that we are looking at the intrinsic properties of the systems here, and not those derived from a set of synthetic transit observations, so the fact that the simulated 1-planet systems do not match the Rayleigh distribution for $\sigma_{e}=0.167$ is not particularly relevant, as we discuss later in this paper when we examine the results of synthetic transit observations of the simulated systems (see section~\ref{sec:syn_obs}).

\begin{figure}
\centering
\includegraphics[width=1.0\columnwidth]{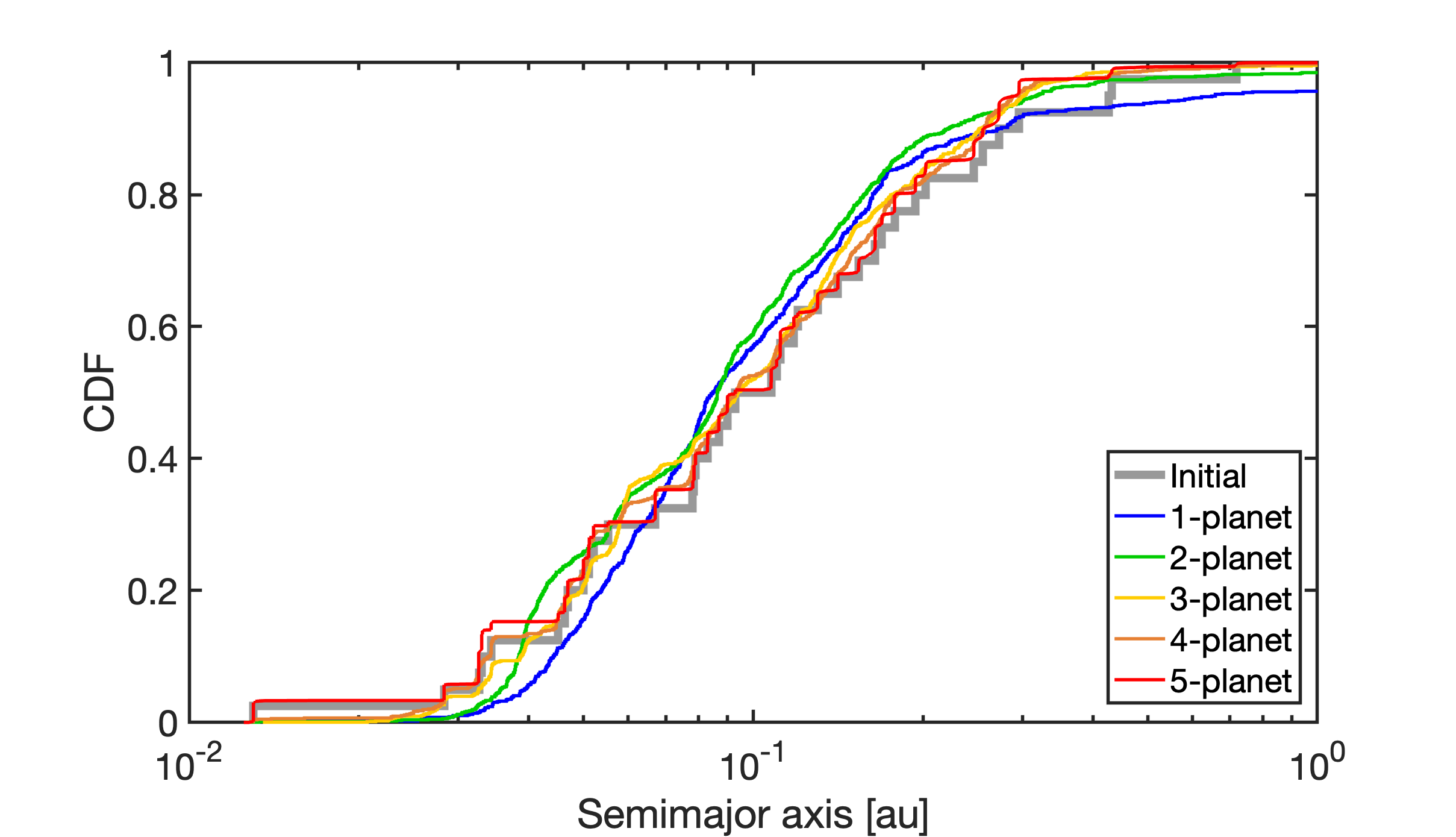}
\caption[Distribution of final semimajor axis]{CDFs of the inner planet semi-major axes obtained from 64 different combinations of inner/outer templates. Different line colours represent different multiplicities. The grey line represents the CDF of the initial semi-major axes.}\label{fig:a_sim}
\end{figure}

Figure \ref{fig:a_sim} shows the cumulative distribution functions for the final semi-major axes of the simulated planets as a function of the final multiplicity. Unlike in figure~\ref{fig:e_sim}, the semi-major axis distributions do not vary strongly with multiplicity, and are very similar to the initial values. The right panel of figure~\ref{fig:inner_e_subplot} shows a pattern of vertical strips indicating that the planets in the unperturbed systems did not move away from their original semi-major axes significantly. Furthermore, we also observe a tendency for the innermost and outermost planets to be the most eccentric for the systems containing 3, 4 and 5 planets. The outermost planets in these systems have experienced the strongest perturbations due to the outer planets, and hence show enhanced eccentricities. The innermost planets that display the largest eccentricities, however, obtained these larger values because of the redistribution of angular momentum deficit \citep[AMD, ][]{2017A&A...605A..72L} within some of the inner planetary systems during the evolution described in Sect.~\ref{subsec:stable}, prior to the runs being performed with the giant planets having been inserted. This point is discussed further in appendix~\ref{app:A1}. For lower multiplicities these patterns becoming increasingly indistinct, which together with the eccentricity distributions mentioned above, indicates that the planets in small multiplicity systems have experienced stronger scattering, as expected.

\subsection{Multiplicities and eccentricities of the outer systems}
The outer planet templates we consider in this study are at or below the RV detection limit (assumed to be 3 ${\rm m \, s^{-1}}$) and are not amenable to transit surveys. Hence, the outcomes of our $N$-body simulations cannot be compared with observations. Previous studies have focused on outer giant systems comprised of planets that are above the RV detection limit \citep[e.g.,][]{2008ApJ...686..580C,2014ApJ...786..101P,2017AJ....153..210H}. \citet{2008ApJ...686..603J} suggested that the dynamical evolution of initially high multiplicity giant systems ($N_{\rm init}>3$ and $m_{\rm p}>0.1$ ${\rm M}_{\rm J} \sim$ $32 ~{\rm M}_{\oplus}$) are likely to result in a lower multiplicity ($N_{\rm final}=\left \{ 2,3 \right \}$). A subset of our simulations with initial conditions lying within those ranges (e.g., \texttt{6g.60M.5AU} and \texttt{6g.100M.10AU}) agree with \citet{2008ApJ...686..603J}, where the final systems with $N_{\rm out,final}=\left \{ 2,3 \right \}$ make up the majority (see right column of table~\ref{tab:out_multi}). 
\begin{table}
	\centering
	\caption{Final multiplicity occurrence rates for a selection of outer system templates (where planets have final $a>1$). Rates in the right-most column are the sum of the occurrence rates for multiplicities of 2 and 3. Note that systems recorded as having zero giant planets actually have surviving planets with $a < 1$.}
	\label{tab:out_multi}
	\begin{tabular}{rrrrrrrr|r} 
		\hline
		\hline
		& \multicolumn{7}{c}{\underline{Multiplicity occurrence rate (\%)}} \\
		Outer template & 0 & 1 & 2 & 3 & 4 & 5 & 6 & $\left \{ 2,3 \right \}$\\ 
		\hline
		\texttt{3g.15M.2AU} 		& 0 & 4 & 82 & 14 & - & - & - & 96\\
		\texttt{3g.30M.2AU} 		& 0 & 16 & 82 & 2 & - & - & - & 84\\
		\texttt{3g.60M.2AU} 		& 2 & 56 & 42 & 0 & - & - & - & 42\\
		\texttt{3g.100M.2AU} 	& 1 & 64 & 54 & 1 & - & - & - & 55\\
		\hline
		\texttt{6g.15M.2AU} 		& 0 & 7 & 30 & 44 & 19 & 0 & 0 & 74\\
		\texttt{6g.30M.2AU} 		& 0 & 18 & 58 & 24 & 0 & 0 & 0 & 82\\
		\texttt{6g.60M.2AU} 		& 0 & 47 & 49 & 4 & 0 & 0 & 0 & 53\\
		\texttt{6g.100M.2AU} 	& 1 & 64 & 34 & 1 & 0 & 0 & 0 & 35\\
		\hline
		\texttt{6g.15M.5AU} 		& 0 & 0 & 0 & 2 & 25 & 56 & 17 & 2\\
		\texttt{6g.30M.5AU} 		& 0 & 2 & 14 & 37 & 44 & 3 & 0 & 51\\
		\texttt{6g.60M.5AU} 		& 0 & 16 & 53 & 30 & 1 & 0 & 0 & 83\\
		\texttt{6g.100M.5AU} 	& 0 & 39 & 53 & 8 & 0 & 0 & 0 & 61\\
		\hline
		\texttt{6g.15M.10AU} 	& 0 & 0 & 0 & 2 & 3 & 38 & 57 & 2\\
		\texttt{6g.30M.10AU} 	& 0 & 0 & 1 & 21 & 37 & 37 & 4 & 22\\
		\texttt{6g.60M.10AU} 	& 0 & 12 & 35 & 40 & 12 & 1 & 0 & 75\\
		\texttt{6g.100M.10AU} 	& 0 & 27 & 61 & 11 & 1 & 0 & 0 & 72\\
		\hline
		\hline
	\end{tabular}
\end{table}
On the other hand, a subset of our simulations do not lie within the parameter space studied by \citet{2008ApJ...686..603J}. Our outer system templates include planets with masses smaller than 0.1 ${\rm M}_{\rm J}$. For those systems (e.g. \texttt{6g.15M.5AU} and \texttt{6g.30M.10AU}), the majority of systems no longer end up with multiplicities of 2 or 3. For example, the template \texttt{6g.15M.5AU} ended up with only 2 \% of systems having 2 or 3 planets. A correlation between the final multiplicities and the masses of the planets can be seen in table~\ref{tab:out_multi}, where the lower mass planets maintain a higher final multiplicity due to less efficient scattering. 

\begin{figure}
\centering
\includegraphics[width=1.0\columnwidth]{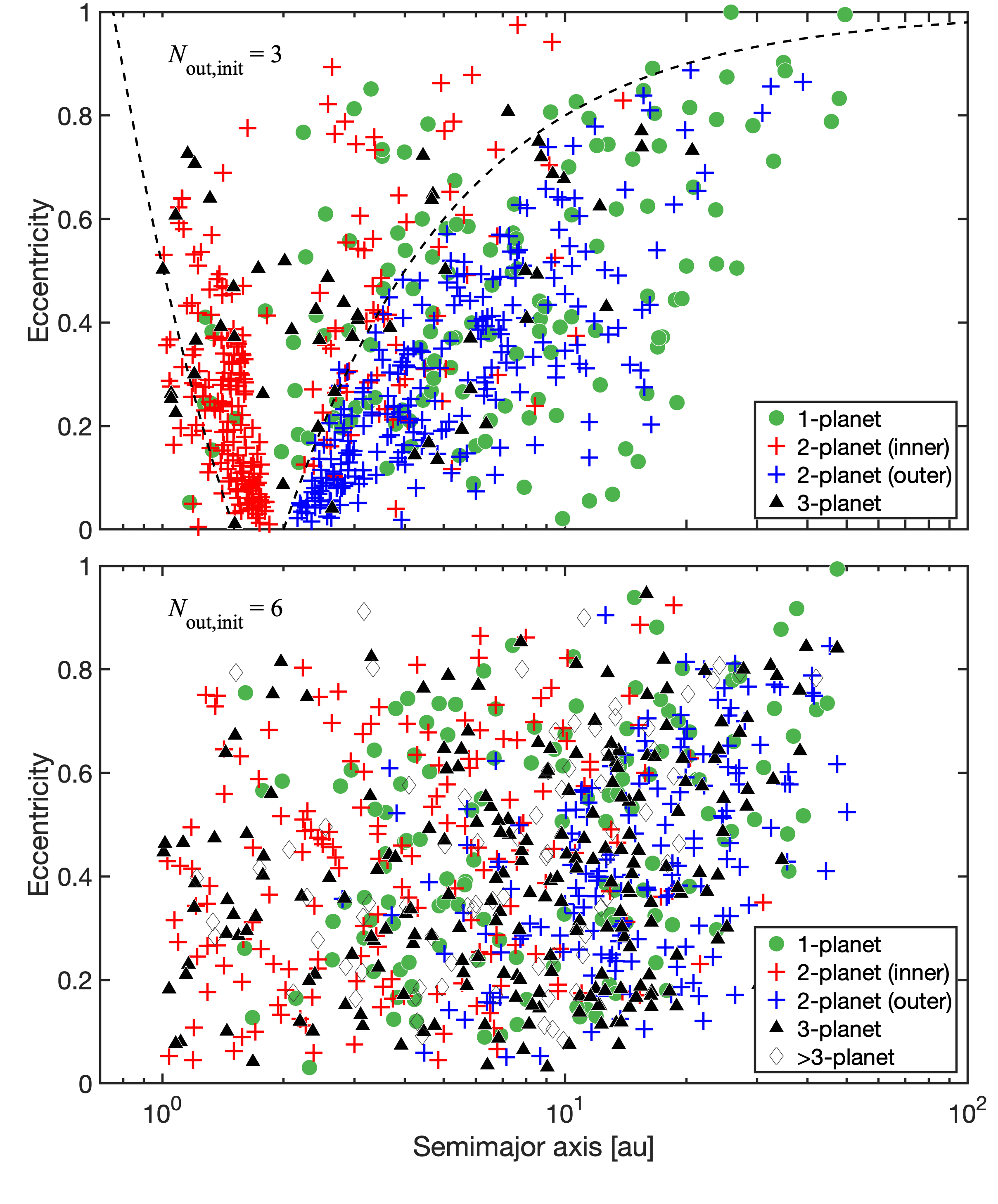}
\caption[Eccentricities and semimajor axis distribution of outer planet draw from \texttt{Kepler55}]{Scatter plots $e$ versus $a$ for the final outer planets resulting from a subset of the \texttt{Kepler55} runs. Different symbols represent the different values of $N_{\rm out,final}$. For the $N_{\rm out,final}=2$ case, the inner planets are marked in red while the outer planets are marked in blue. The upper panel shows results for systems where the outer systems had $N_{\rm out,init}=3$ planets and $\tilde{a}_{\rm out,init}=2{\rm~au}$. The left dashed line denotes the value of $a=1.51/(1+e)$ and the right dashed line denotes the value $a=2.0/(1-e)$. The bottom panel shows the $e$ versus $a$ distribution from a subset of outer templates with $N_{\rm out,init}=6$ and $\tilde{a}_{\rm out,init}=2{\rm~au}$.}\label{fig:55_e_2au}
\end{figure}

The upper panel of figure~\ref{fig:55_e_2au} shows the eccentricities versus semi-major axes of the outer giant planets from the \texttt{Kepler55.3g} ($N_{\rm out,init}=3$) runs. According to table~\ref{tab:out_multi}, most of the systems in this subset end up with $N_{\rm out,final}=2$. This agrees with previous studies that considered three outer planets, where $N_{\rm out,final}=2$ is the most common outcome \citep[e.g.,][]{2014ApJ...786..101P,2017AJ....153..210H}. Furthermore, the 2-planet systems show a v-shaped distribution, which can also be seen in the simulation results of \citet{2008ApJ...686..580C}, \citet{2014ApJ...786..101P}, and \citet{2017AJ....153..210H}. Similar to \citet{2008ApJ...686..580C}, we demonstrate that there exists a separation between the inner and outer planets in the population of the 2-planet systems. A clear division between the two sub populations can be seen, where the inner planet tends to follow the apoapsis, $a=a_{\rm in}/(1+e)$, of the initial innermost planet (left dashed line in figure~\ref{fig:55_e_2au}), and the outer planet tends to follow the periapsis, $a=a_{\rm mid}/(1-e)$, of the middle planet of the three original planets (right dashed line). The $a_{\rm in}$ and $a_{\rm mid}$ that applied to the dashed lines in figure \ref{fig:55_e_2au} are 1.51 and 2.00~au respectively, where this value came from the semi-major axes of the innermost and middle planets in the outer systems that make up the \texttt{Kepler55.3g.100M.2AU} template.

\begin{figure}
\centering
\includegraphics[width=1.0\columnwidth]{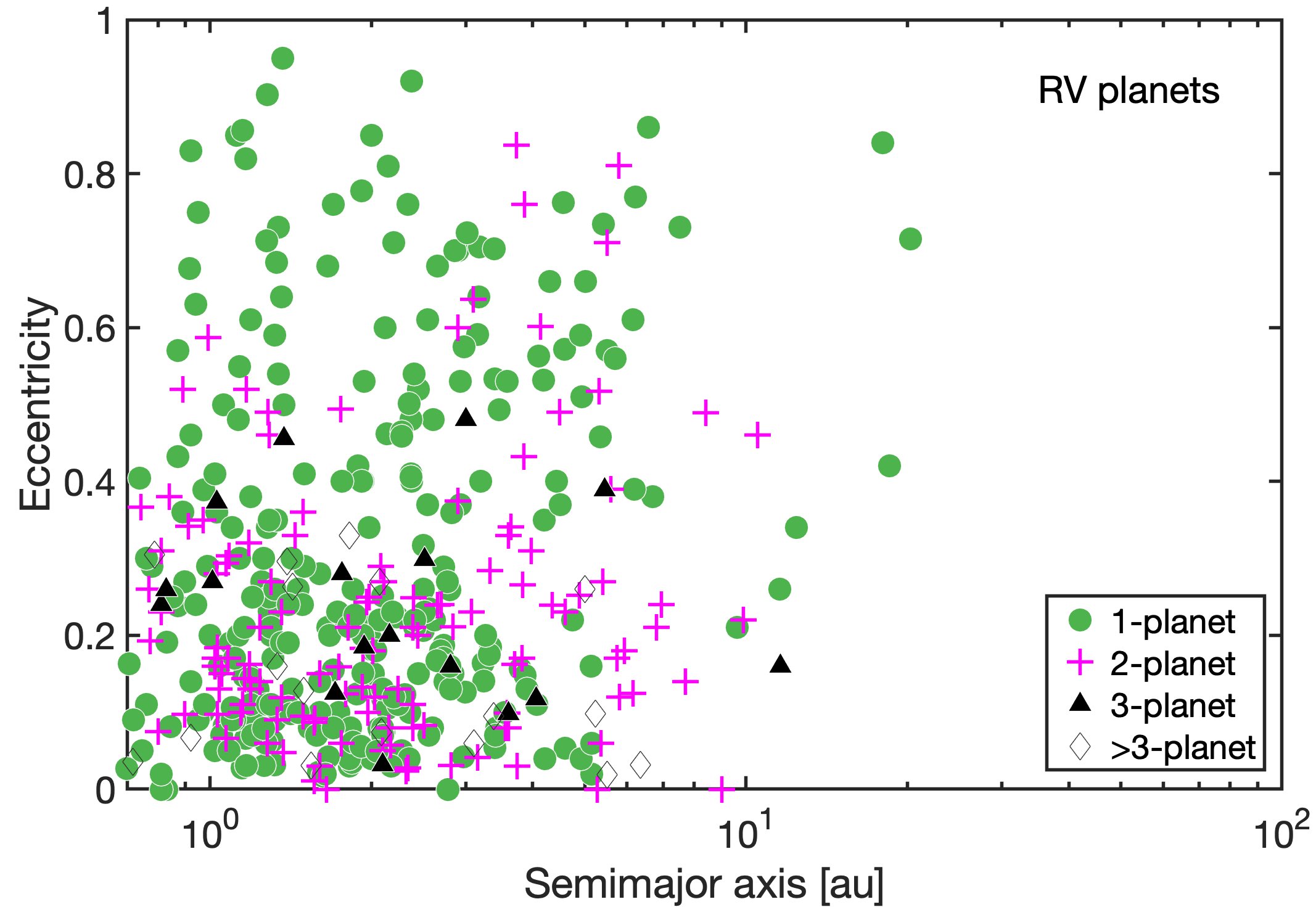}
\caption[Eccentricities and semimajor axis of RV planets]{Eccentricities versus semi-major axes of the observed RV planets. 1-planet systems are marked as green circles, 2-planet systems are marked as pink crosses, 3-planet systems are marked as black triangles, and RV systems with multiplicities higher than three are marked with unfilled black diamonds.} \label{fig:ae_RV}
\end{figure}

The v-shaped distribution seen in the upper panel cannot be seen in the results from the higher initial multiplicity templates, such as \texttt{Kepler55.6g.2AU} (figure~\ref{fig:55_e_2au}, bottom panel), because the larger amount of scattering washes this feature out. This feature can also not be seen in the $e$ versus $a$ plot for the observed RV planets shown in figure~\ref{fig:ae_RV}, indicating that this population is not consistent with it being the result of scattering from initial conditions similar to those in the \texttt{Kepler55.3g} ($N_{\rm out,init}=3$) runs.

\section{Synthetic observation of the final planetary systems}\label{sec:syn_obs}
The multiplicity of a planetary system observed using the transit method depends on the viewing angle, intrinsic multiplicity and mutual inclinations between the planets within the system, the radius of the host star and orbital radii of the planets (where we have ignored the effect of the finite planet radius). To compare the outcomes of our $N$-body simulations with analyses of Kepler data relating to the multiplicity and eccentricity dichotomies, we now present the results of synthetically observing the simulation results.

Each simulated planetary system is synthetically observed from 10,000 randomly chosen viewing locations, isotropically distributed with respect to each host star. We only consider planets that are within the inner planetary system ($a_{\rm p}<1$ au). As the smallest planet we considered in the simulations (Kepler-62c) is a confirmed Kepler planet, we assume all planets satisfy the observation limits of a Kepler-like survey.

\subsection{Observed multiplicities}
Following the approach of \citet{2012ApJ...758...39J}, we consider the relative numbers of one-planet, two-planet, ..., five-planet systems detected when the simulation outcomes are synthetically observed. Similar to \citet{2020MNRAS.491.5595P}, using the observed numbers of one-planet, two-planet, etc. systems, we then define a Transit Multiplicity Ratio (abbreviated to TMR hereafter) as follows:
\begin{equation}\label{eq:transitratio}
\textrm{TMR}(i:j)=\frac{\textrm{Number of } i\textrm{-planet systems}}{\textrm{Number of }j\textrm{-planet systems}},
\end{equation}
where $i$ and $j$ represent the numbers of planets detected during each synthetic observation of each system.

\begin{figure}
\centering
\includegraphics[width=1.0\columnwidth]{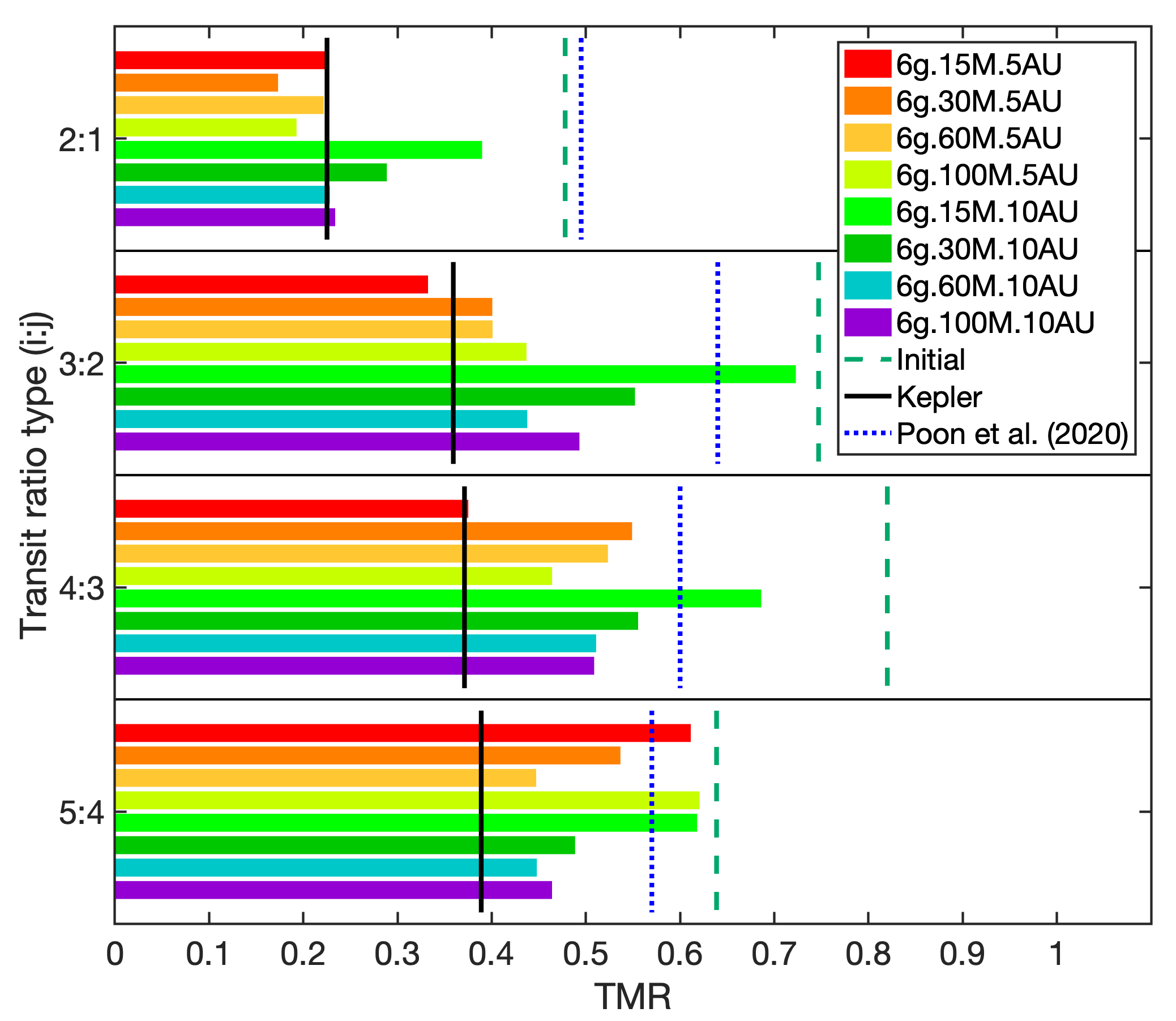}
\caption[Transit multiplicity ratio]{Synthetic TMRs from the simulations. Different coloured bars denote TMRs of different outer templates. The green dashed vertical lines show the TMRs of the initial inner systems, the black vertical lines show the observed Kepler TMRs, and the blue dotted lines show the TMRs obtained from $N$-body simulations of in situ formation of super-Earths by \citet{2020MNRAS.491.5595P}.}\label{fig:transitratio}
\end{figure}

The TMR values obtained are shown in figure~\ref{fig:transitratio}. The coloured histograms show the values obtained for each of the outer system templates, the jade vertical dashed lines mark the TMR values of the initial conditions, the black vertical lines mark the TMRs from the Kepler data, and the blue vertical dotted line shows the results obtained by \citet{2020MNRAS.491.5595P} from their simulations of in situ formation of super-Earth systems. 

The TMRs drop below the initial values for all outer templates and for all multiplicity ratios. For the 2:1 and 3:2 TMRs, the systems with $\tilde{a}_{\rm out}=5$~au provide somewhat better fits to the Kepler TMRs than those with $\tilde{a}_{\rm out}=10$~au. Considering the 2:1 TMR panel in particular, the \texttt{6g.15M.5AU}, \texttt{6g.60M.5AU}, and \texttt{6g.60M.10AU} templates have an almost exact match to the Kepler 2:1 TMR. Very close agreement is obtained between the Kepler 3:2 TMR and those produced by the templates \texttt{6g.15M.5AU}, \texttt{6g.30M.5AU} and \texttt{6g.60M.5AU}. Apart from the \texttt{6g.15M.5AU} template, however, the 3:2 TMRs obtained from the synthetic observations are higher than obtained from the Kepler data, indicating that the simulations are over-producing systems of 3 transiting planets relative to 2 transiting planets. Indeed, for higher multiplicity ratios than 2:1 more generally, the simulated TMRs are higher than the corresponding Kepler TMRs, suggesting that the mutual inclinations within the simulated systems are smaller than among the Kepler systems. 

When considering the TMRs across all multiplicity ratios, we see that the template \texttt{6g.15M.10AU} is consistently a factor of $\sim$1.5-2 above the corresponding Kepler TMRs, and can be considered the worst performing template. The TMRs for this template remain closer to the initial values compared to the others, because the majority of final systems for this template are unperturbed systems (see figure~\ref{fig:planet_no_all_template}), so it is expected that the change in TMRs would be limited. While no template fits the Kepler TMRs for all multiplicity ratios, the best performing overall are \texttt{6g.15M.5AU}, \texttt{6g.60M.5AU}, \texttt{6g.100M.5AU} and \texttt{6g.60M.10AU}. Each of these provides a decent fit to the Kepler TMRs for three of the multiplicity ratios considered in figure~\ref{fig:transitratio}, and each performs relatively poorly in one of multiplicity ratios.

In figure~\ref{fig:transitratio}, comparison can also be made between the simulated TMRs to the TMRs from the in situ formation simulations in \citet{2020MNRAS.491.5595P}, where the mutual inclinations of the super-Earths are self-excited by gravitational scattering between the planets as they form. In general (except for the 3:2 TMR for \texttt{6g.15M.10AU}), the TMRs obtained in this current study are in significantly closer agreement with the Kepler observations than those obtained for systems that do not experience perturbations from outer planets.

\subsection{Eccentricity distributions}
In a recent study, \citet{2016PNAS..11311431X} suggested that the mean eccentricity of Kepler single planet systems is significantly higher ($\langle e \rangle\approx0.3$) than that of the multi-planet systems ($\langle e \rangle\approx0.04$). More recently, \citet{2019AJ....157..198M} also showed there is an eccentricity dichotomy, and they obtained $\langle e \rangle\approx0.21$ for singles and $\langle e \rangle\approx0.05$ for multi-planet systems. 

Recent studies that consider the dynamics of planets in compact inner systems show that a small dichotomy arises between the eccentricities of the single and multi-transiting systems, but these fail to reproduce the large eccentricity values for the 1-planet systems \citep[e.g.][]{2020MNRAS.491.5595P,2020ApJ...891...20M}. An increase in the orbital eccentricity of a planet can be induced by planet-planet scattering, but this is limited to producing values $e_{\rm p} \sim v_{\rm e}/v_{\rm K}$, where $v_{\rm e}/v_{\rm K}$ represent the escape velocity from a planet and the Keplerian velocity. \citet{2019AJ....157...61V} and \citet{2020MNRAS.491.5595P} argued that higher mass components are needed in order for scattering to produce systems with high enough eccentricities to match the observations.

Table~\ref{tab:obs_ecc} lists the mean and median eccentricities from our synthetic transit observations. Each template shows a clear eccentricity dichotomy, where the values of the means and medians of the eccentricities for single-transiting systems are always larger than the values for multi-transiting systems ($\{\langle e_{1} \rangle,\tilde{e}_{1}\}>\{\langle e_{\geq 2} \rangle,\tilde{e}_{\geq 2}\}$). The dichotomy signal becomes stronger as the masses of the outer planets increases and the initial orbital radii decrease. For example, the \texttt{6g.100M.5AU} template gives $\{\langle e_{1} \rangle,\langle e_{\geq 2} \rangle\}=\{0.25,0.08\}$, while the template with lower mass, \texttt{6g.15M.5AU}, gives $\{\langle e_{1} \rangle,\langle e_{\geq 2} \rangle\}=\{0.15,0.07\}$. The template \texttt{6g.15M.10AU} has the lowest mass and the most distant outer planets, and only displays a small eccentricity dichotomy signal ($\{\langle e_{1} \rangle,\langle e_{\geq 2} \rangle\}=\{0.11,0.06\}$). The value of $\{\langle e_{1} \rangle,\langle e_{\geq 2} \rangle\}$ for the template \texttt{6g.15M.10AU} is only slightly higher than the value for the control set of simulations (only the inner Kepler templates and no outer systems), as the multiplicities shown in figure~\ref{fig:planet_no_all_template} illustrate that $\sim75$ \% of the runs using this template resulted in an unperturbed system. 

\begin{table}
	\centering
	\caption{Mean (upper table) and median (lower table) eccentricity of the inner systems obtained by the synthetic transit observations in different outer templates. The observed mean eccentricities are listed by their observed multiplicity. The subscript 1 represents the single-transit systems and subscript $\geq2$ represents the multi-transiting systems. The control set did not include outer systems in the runs (see section~\ref{subsec:stable}).}
	\label{tab:obs_ecc}
	\begin{tabular}{rrrrrrr} 
		\hline
		\hline
		& \multicolumn{6}{c}{\underline{Mean eccentricity}} \\
		Template & $\langle e_{1} \rangle$ & $\langle e_{2} \rangle$ & $\langle e_{3} \rangle$ & $\langle e_{4} \rangle$ & $\langle e_{5} \rangle$ & $\langle e_{\geq 2} \rangle$\\ 
		\hline
\texttt{6g.15M.5AU}   & 0.21 & 0.10 & 0.07 & 0.05 & 0.06 & 0.09 \\
\texttt{6g.30M.5AU}   & 0.23 & 0.09 & 0.05 & 0.05 & 0.06 & 0.09 \\
\texttt{6g.60M.5AU}   & 0.25 & 0.09 & 0.06 & 0.05 & 0.05 & 0.07 \\
\texttt{6g.100M.5AU}  & 0.25 & 0.08 & 0.06 & 0.04 & 0.05 & 0.08 \\
\texttt{6g.15M.10AU}  & 0.11 & 0.06 & 0.05 & 0.04 & 0.05 & 0.06 \\
\texttt{6g.30M.10AU}  & 0.17 & 0.08 & 0.06 & 0.05 & 0.05 & 0.08 \\
\texttt{6g.60M.10AU}  & 0.23 & 0.11 & 0.07 & 0.05 & 0.06 & 0.10 \\
\texttt{6g.100M.10AU} & 0.22 & 0.11 & 0.06 & 0.05 & 0.04 & 0.08 \\
Control				 & 0.08 & 0.05 & 0.04 & 0.04 & 0.04 & 0.05 \\
\hline
\\
& \multicolumn{6}{c}{\underline{Median eccentricity}} \\
		Template & $\tilde{e}_{1}$ & $\tilde{e}_{2}$ & $\tilde{e}_{3}$ & $\tilde{e}_{4}$ & $\tilde{e}_{5}$ & $\tilde{e}_{\geq 2}$\\ 
\hline
\texttt{6g.15M.5AU}   & 0.15 & 0.08 & 0.06 & 0.04 & 0.06 & 0.07 \\
\texttt{6g.30M.5AU}   & 0.17 & 0.07 & 0.04 & 0.04 & 0.06 & 0.07 \\
\texttt{6g.60M.5AU}   & 0.20 & 0.07 & 0.05 & 0.04 & 0.04 & 0.06 \\
\texttt{6g.100M.5AU}  & 0.23 & 0.07 & 0.05 & 0.04 & 0.05 & 0.06 \\
\texttt{6g.15M.10AU}  & 0.08 & 0.05 & 0.04 & 0.04 & 0.04 & 0.05 \\
\texttt{6g.30M.10AU}  & 0.11 & 0.06 & 0.05 & 0.04 & 0.05 & 0.06 \\
\texttt{6g.60M.10AU}  & 0.16 & 0.09 & 0.06 & 0.05 & 0.06 & 0.07 \\
\texttt{6g.100M.10AU} & 0.16 & 0.07 & 0.05 & 0.04 & 0.04 & 0.06 \\
Control		      	 & 0.07 & 0.05 & 0.04 & 0.04 & 0.04 & 0.05 \\
		\hline
		\hline
	\end{tabular}
\end{table}

\begin{figure}
\centering
\includegraphics[width=1.0\columnwidth]{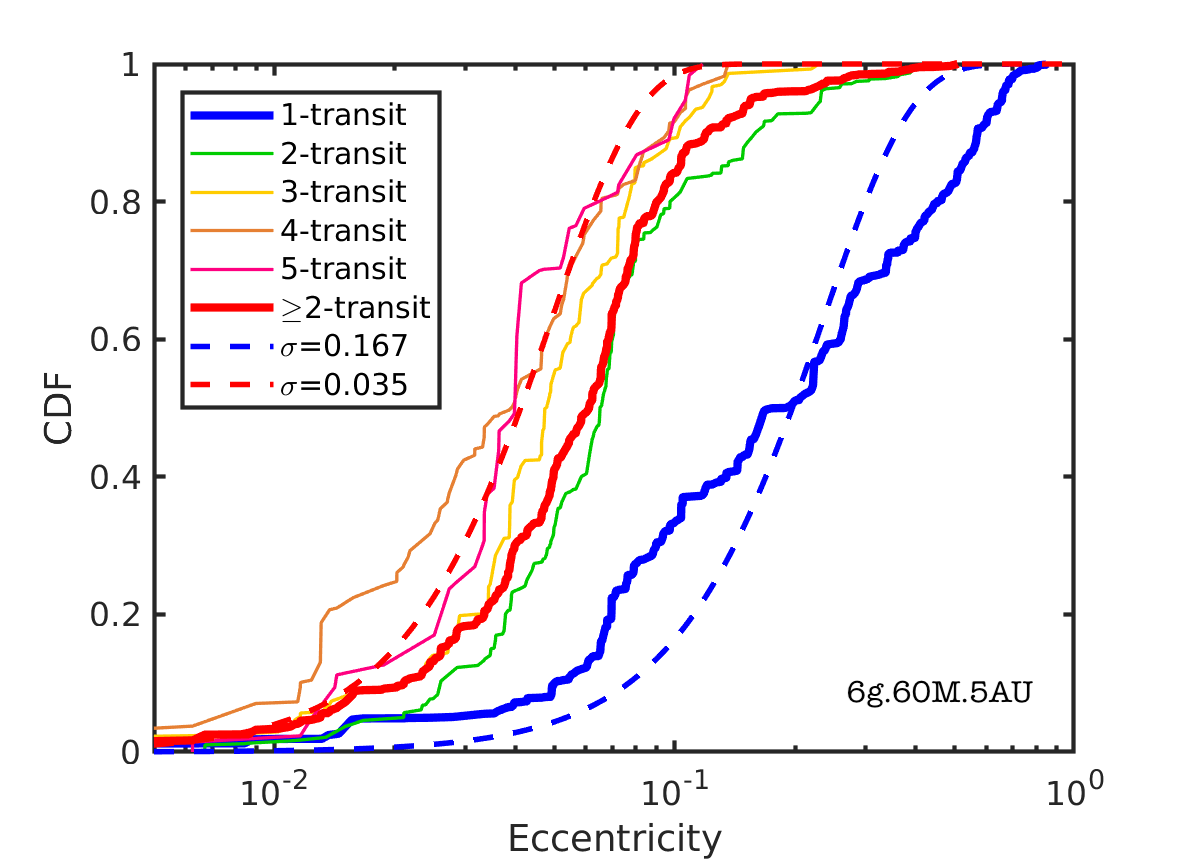}
\caption[Synthetic observed eccentricities CDFs for \texttt{6g.60M.5AU}]{CDFs of the eccentricities obtained from the synthetic observations of the simulations for the \texttt{6g.60M.5AU} template. For comparison, the CDFs of eccentricities drawn from Rayleigh distributions with eccentricity parameters $\sigma=0.167$ and $\sigma=0.035$ are also plotted.}\label{fig:e_all60M5AU}
\end{figure}

Rather than just considering the mean and median values of the eccentricity distributions, \citet{2019AJ....157..198M} supposed the distributions follow a Rayleigh distribution and concluded that under this assumption the eccentricity parameters $\sigma_{e}=0.167$ and 0.035 gave the best fits to the Kepler single and multiple systems, respectively. Figure~\ref{fig:e_all60M5AU} shows the eccentricity distributions of the synthetically observed single and multiple systems (thick solid blue and red lines) from the \texttt{6g.60M.5AU} template, together with the Rayleigh distributions suggested for the Kepler single and multiple systems (dashed blue and red lines) \citep{2019AJ....157..198M}. Similar plots for all eight outer system templates are shown in figures~\ref{fig:CDF5AU} and \ref{fig:CDF10AU} in the appendix, and although they differ in detail the plots show similar behaviours and trends. In figure~\ref{fig:e_all60M5AU} we see that the synthetically observed $e$-distribution for the single-transiting systems does not follow a Rayleigh distribution very closely, but nonetheless it has a median eccentricity very similar to that of the suggested Rayleigh distribution with $\sigma_{e}=0.167$. For the multiple systems, their distributions are closer to a Rayleigh distribution, and we see that as the multiplicity decreases there is a tendency for the eccentricities to increase. For example, the 2-planet systems produce a more eccentric distribution than the 4- or 5-planet systems, and this is because on average the lower multiplicity systems come from observations of systems that have been more strongly perturbed by the outer planets. Hence, the synthetically observed systems show the same inverse relation between eccentricity and observed multiplicity that was noted in section~\ref{subsubset:no_and_e_inner} when discussing the intrinsic properties of the simulated planetary systems, albeit at a lower level of significance.

Figure~\ref{fig:e_all60M5AU} also shows that the 2-planet systems (yellow line) provide the main contribution to the eccentricity distribution for all multi-planet systems (thick red line), and the effect of this is to shift the eccentricity distribution away from that displayed by the control set that characterises the dynamics of the unperturbed inner systems, to one corresponding to systems that are more dynamically excited. This effect reduces the dichotomy signal between the single and multiple systems that comes out of the synthetic observations because the detected 2-planet systems are generally more excited than the higher multiplicity systems. 

Considering both the TMRs and the mean eccentricities together, the template \texttt{6g.60M.5AU} appears to be the best performing overall. The mean eccentricities for single and multiple systems show a strong dichotomy $\{\langle e_{1} \rangle,\langle e_{\geq 2} \rangle\}=\{0.25,0.07\}$ that is in decent agreement with the Kepler eccentricity dichotomy $\{\langle e_{1} \rangle,\langle e_{\geq 2} \rangle\}=\{0.3,0.04\}$ \citep{2016PNAS..11311431X} or 
$\{\langle e_{1} \rangle,\langle e_{\geq 2} \rangle\}=\{0.21,0.05\}$ \citep{2019AJ....157..198M}. The 2:1, 3:2 and 5:4 TMRs are in very good agreement with the Kepler TMRs, with only the 4:3 TMR being in significant disagreement. The template \texttt{6g.100M.5AU} also performs well, while for the more distant systems of outer planets centred around 10 au the template \texttt{6g.60M.10AU} is the best performing.

\section{Impact of additional physics}
\label{sec:unmodel}
In this section we consider the impact of different physical processes that were not included in the main suite of simulations presented in earlier sections.
\subsection{Relativistic precession}
Precession due to General Relativity (GR) can become a significant effect for planets orbiting close to their host stars, and in particular can influence the secular interactions within planetary systems, as shown for example in the recent study by \cite{2020MNRAS.493..427M}. As discussed in section~\ref{subsubsec:dynamic}, the dominant effect that perturbs the inner systems in our study is the mutual scattering of outer giants, such that during pericentre passage they enter the inner system and cause strong scattering of the planets there. We do not expect GR to provide a stabilising effect in this situation, and this expectation appears to be confirmed by \cite{2017AJ....153..210H}, who examined the influence of GR in their study of planetary scattering. 

\subsection{Planet-planet collisions}
\label{subset:collision}
Our simulations adopted a perfect merger treatment of planet-planet collisions, and a more realistic prescription might lead to removal of mass from the colliding bodies \citep{2012ApJ...745...79L,2012ApJ...751...32S}, especially for high-velocity close-in and eccentric collisions. To investigate the changes to our main results that arise when using a more realistic collision model, we have rerun a sub-set of simulation using a version of \textsc{symba} \citep{1998AJ....116.2067D,2020MNRAS.491.5595P,2020MNRAS.493.4910S}, which implements the collision algorithm from \citet{2012ApJ...745...79L}. Here, the outcome of a collision falls into one of nine regimes: supercatastrophic disruption, catastrophic disruption, erosion, partial accretion, hit-and-spray, hit-and-run, bouncing collision, graze-and-merge, and perfect merger, depending on the collision conditions. The total mass before and after a collision is conserved and obeys the relation of,
\begin{equation}
M_{\mathrm{Total}}=M_{\mathrm{LR}}+M_{\mathrm{SLR}}+M_{\mathrm{Total,debris}},
\end{equation}
where $M_{\mathrm{Total}}$ is the total mass of the colliding bodies before the collision, $M_{\mathrm{LR}}$ is the mass of the post-collision largest remnant, $M_{\mathrm{SLR}}$ is the mass of the second largest remnant, and $M_{\mathrm{Total,debris}}$ is the total mass of the debris particles that generated during the collision. For a detailed description of the adopted version of \textsc{symba}, we refer the reader to the model descriptions in \citet{2020MNRAS.491.5595P} and \citet{2020MNRAS.493.4910S}.

\begin{figure}
\includegraphics[width=1.0\columnwidth]{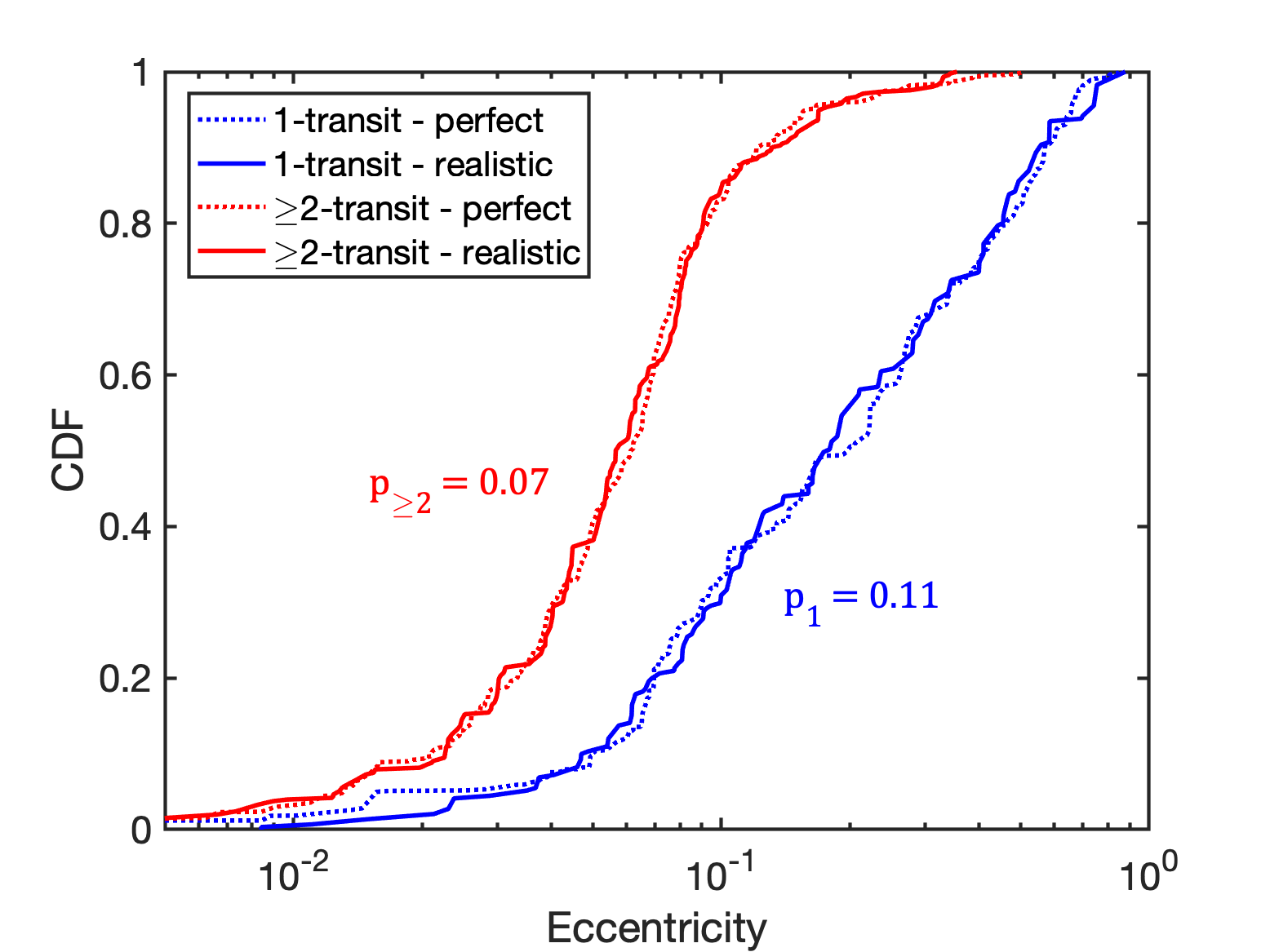}
\caption[Perfect vs. realistic collisions]{CDFs of the eccentricities obtained from synthetic observations of the simulation outcomes for the \texttt{6g.60M.5AU} template by the perfect (dotted lines) and the realistic (solid lines) accretion model. The dotted lines are the same as the thick solid lines from figure~\ref{fig:e_all60M5AU} in the same colours. The $p$-values from the K-S test are $p_{1}=0.11$ and $p_{\geq2}=0.07$. }\label{fig:pvr}
\end{figure}

We consider the runs from the subset labelled \texttt{6g.60M.5AU} (8 inner planetary systems $\times$ 100 runs each $=800$ runs in total). Given we are using \textsc{symba} and not \textsc{mercury} for these simulations, we ran the same 800 set of initial conditions using both perfect accretion and the more realistic collision model. Figure~\ref{fig:pvr} shows the cumulative distribution function of eccentricities obtained from the synthetic transit observation of the final systems. The different collision models provide similar distributions of the observed eccentricities for both single- and multi-transit systems.  They lead to similar mean and median values, where the perfect accretion models have $\lbrace\langle e_{1} \rangle,\tilde{e}_{1},\langle e_{\geq2} \rangle,\tilde{e}_{\geq2}\rbrace=\lbrace 0.25,0.20,0.07,0.06\rbrace$ (table \ref{tab:obs_ecc}) and realistic collision models have $\lbrace 0.25,0.18,0.07,0.06\rbrace$. Moreover, the Kolmogorov-Smirnov test (K-S test) between the two sets of synthetically observed eccentricities yields $p$-values for the single- and multi-transit of $p_{1}=0.11$ and $p_{\geq2}=0.07$, respectively. \citet{2020MNRAS.491.5595P} demonstrates that the re-accretion of the collision debris occurs quickly (on a time-scale of $10^{3}$~yr) and most of the debris (>80 \%) is re-accreted back by the largest/second largest remnant. This test shows that our results are insensitive to the collision prescription used in the simulations.

\subsection{Tidal dissipation}
\label{subset:tidal}
\begin{figure}
\includegraphics[width=1.0\columnwidth]{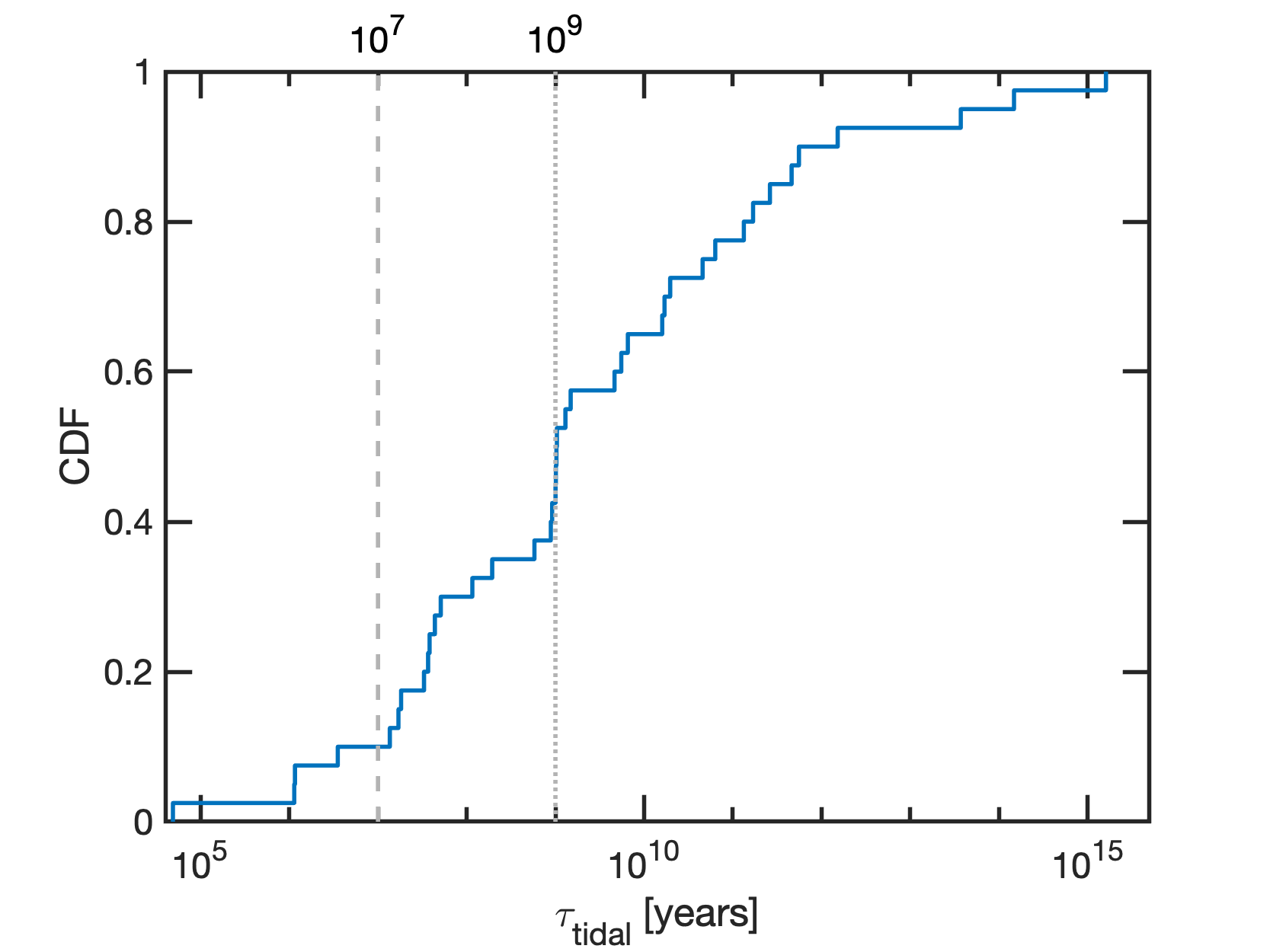}
\caption[Tidal time-scale of the inner system templates]{CDFs of the tidal eccentricity damping time-scale obtained by equation \ref{eq:tidaltime}. The vertical dashed line marked the 10 Myr time (main simulation runtime) and the dotted line marked the 1 Gyr time.}\label{fig:tidtime}
\end{figure}

The simulations presented in this study considers the dynamics of close-in super-Earths, for which tidal interactions with the central star could lead to significant eccentricity damping. \citet{1966Icar....5..375G} \citep[see also][]{2008ApJ...678.1396J} derived the tidal eccentricity damping timescale of the planet, $\tau_{\mathrm{tidal}}$ as
\begin{equation}\label{eq:tidaltime}
\tau_{\mathrm{tidal}}= \left [\dfrac{4}{63}\left (GM^3_{\star}\right )^{-1/2} \dfrac{M_{\mathrm{p}}}{R_{\mathrm{p}}^{5}} Q_{\mathrm{p}}   \right ]a_{\mathrm{p}}^{13/2},
\end{equation}
where $R_{\mathrm{p}}$ is the radius of the planet and $Q_{\mathrm{p}}$ is the tidal dissipation parameter. Figure~\ref{fig:tidtime} shows the values of $\tau_{\mathrm{tidal}}$ obtained from equation~\ref{eq:tidaltime} with the value of $Q_{\mathrm{p}}=100$ for all the planets contained in the inner planetary templates. The appropriate value of $Q_{\mathrm{p}}$ is uncertain, and is thought to range between $100 \lesssim Q_{\mathrm{p}} \lesssim 10^4$, with smaller values applying to Earth-like bodies and the larger values applying to planets with significant gas envelopes. For the adopted value of $Q_{\mathrm{p}}$ we see that $10~\%$ of the planets in the inner systems have $\tau_{\mathrm{tidal}}$ smaller than the simulation runtime of $10^7$ yr, and more importantly $\sim 50\%$ of the planets have $\tau_{\mathrm{tidal}}\le 1$~Gyr, indicating that tides should play an important role over the typical ages of the Kepler systems.

Equation~\ref{eq:tidaltime} shows that the eccentricity damping time is a strong function of semi-major axis, and hence within a multiplanet system we would expect the inner-most planet to experience the strongest tidal damping. Secular interactions combined with tidal dissipation, however, can increase the efficiency with which the eccentricities of more distantly orbiting planets are damped, and hence it is necessary to consider the coupled evolution of entire planetary systems when considering the effects of tidal dissipation on the observed eccentricity distribution. To examine the effect of tides on our results, we have extended the simulations labelled as \texttt{6g.60M.5AU}, including tidal forces on all planets operating on timescales given by equation~\ref{eq:tidaltime}. Ideally we would run the simulations for 1~Gyr, but this is not possible for the short-period systems we are considering, so we instead adopt the value of $Q_{\mathrm{p}}=1$ and a runtime of 10 Myr, which should be equivalent to running the simulations for 1 Gyr with $Q_{\mathrm{p}}=100$ because of the linear relationship between $Q_{\mathrm{p}}$ and $\tau_{\mathrm{tidal}}$ .

\begin{figure}
\includegraphics[width=1.0\columnwidth]{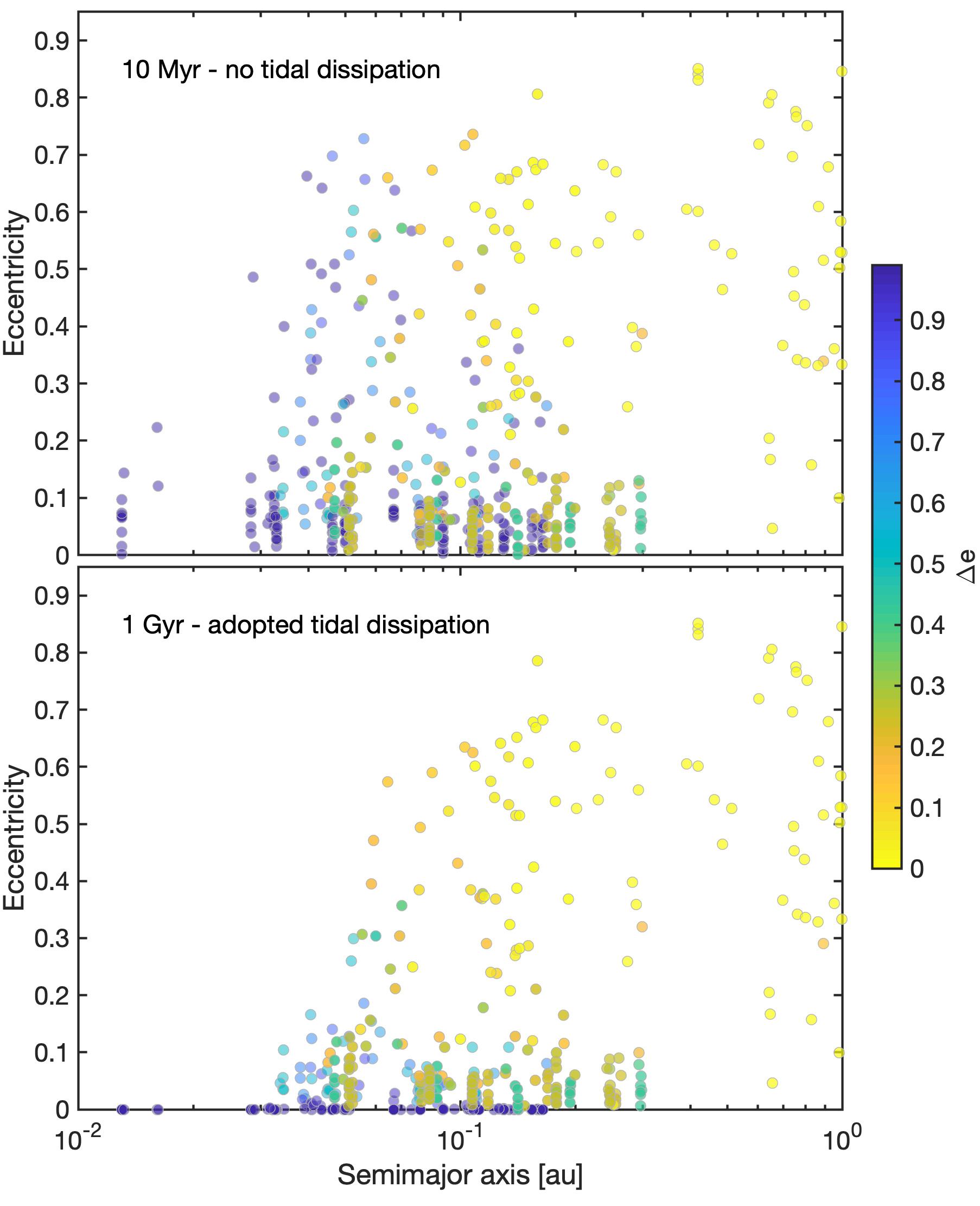}
\caption[No tidal vs. tidal]{Eccentricities versus semi-major axes of the simulation results from \texttt{6g.60M.5AU} template. Upper panel: the simulation results from the main simulations (10 Myr $N$-body, without tidal dissipation); Lower panel: the simulation results of the next 10 Myr $N$-body $+$ 1 Gyr adopted tidal evolution. The colour for each data point illustrates the fractional change of eccentricity, $\Delta e$.}\label{fig:tideff}
\end{figure}

Figure~\ref{fig:tideff} shows the eccentricity distributions for the \texttt{6g.60M.5AU} subset of runs, where the upper panel shows the results after the first 10 Myr without tidal forces applied and the lower panel shows the results after another 10 Myr during which tidal damping was applied. The colour for each data point illustrates the fractional change of eccentricity, $\Delta e= (e_{\rm i} - e_{\rm f})/e_{\rm i}$. As expected, tidal damping is more effective for close-in planets ($a\lesssim0.1~\mathrm{au}$). The heavily damped planets (blue data points) contain $\sim30-40~\%$ of the overall population, which is similar to the CDF value at $10^9$ yr from Figure~\ref{fig:tidtime}. These heavily damped planets end up in essentially circular orbits. 

\begin{figure}
\includegraphics[width=1.0\columnwidth]{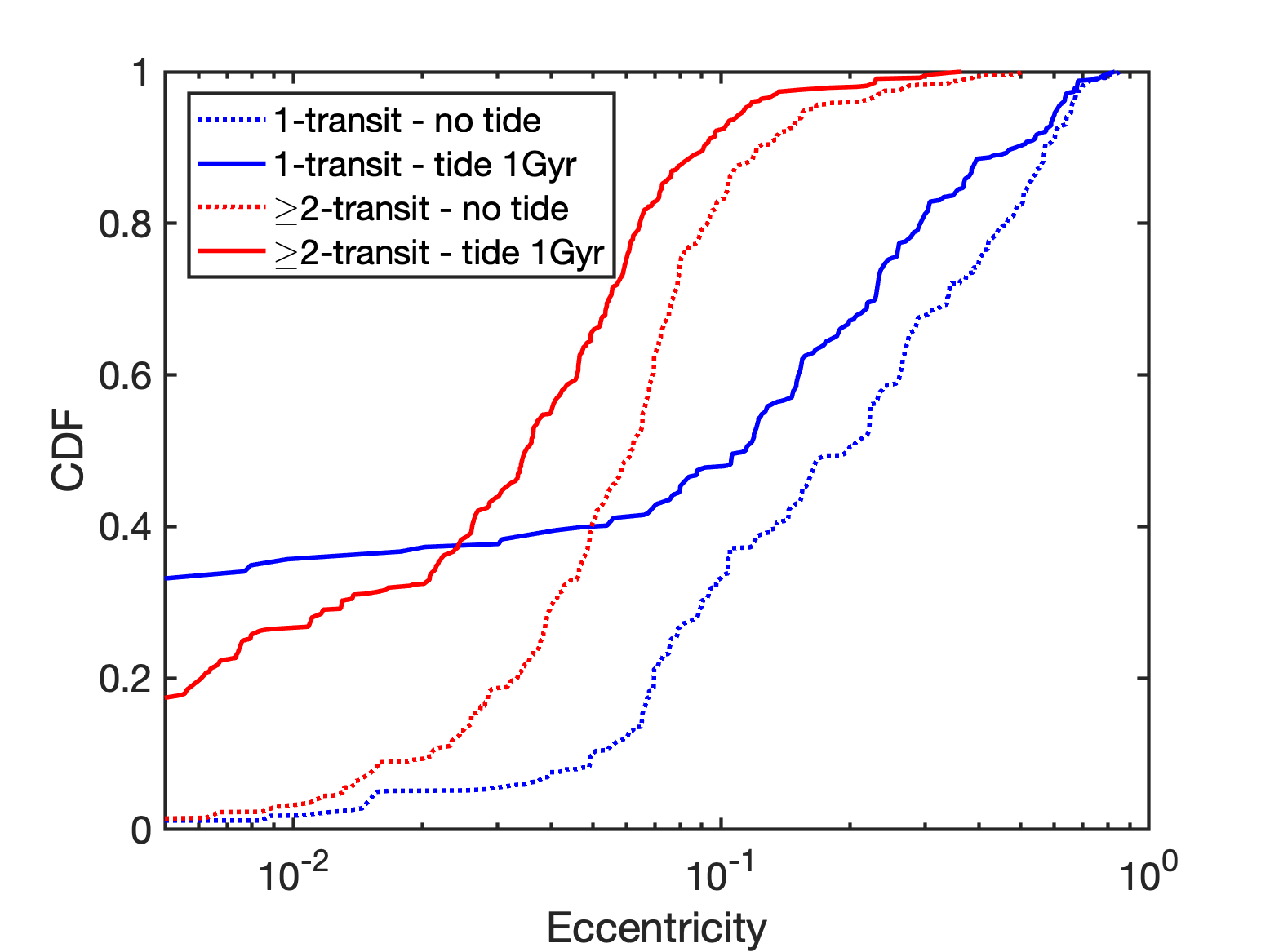}
\caption[Tidal vs. no tidal damping]{CDFs of the eccentricities obtained from the synthetic observations of the simulations for the \texttt{6g.60M.5AU} template by the 10 Myr no tidal damping model (dotted lines) and the next 10 Myr $+$ 1 Gyr tidal damping model (solid lines). The dotted lines are the same as the thick solid lines from figure \ref{fig:e_all60M5AU} in the same colours.}\label{fig:eidcom}
\end{figure}

Figure~\ref{fig:eidcom} compares the synthetically observed eccentricity distributions for the systems with and without tidal damping. Around 35~\% of the single-transiting planets and 20~\% of the planets in multi-transiting systems end up with very low eccentricities ($e<0.01$), and we see in general that tides cause a significant shift in the observed eccentricity distributions of both single- and multi-transiting systems. A contributing factor in explaining these changes is the fact that the most heavily damped planets are also the innermost planets of the systems, which have the highest probability to transit ($P \sim R_{\star}/a_{\mathrm{p}}$). However, more distantly orbiting planets are also observed, and these are not strongly affected by tides, and so contribute a significant number of high eccentricity planets to the observed distributions. It is interesting to note that a clear eccentricity dichotomy is maintained, even in the presence of eccentricity damping, when comparing the eccentricities of single- and multi-transiting planets.

The shifted eccentricity distributions have mean and median eccentricities of $\lbrace\langle e_{1} \rangle,\tilde{e}_{1}\rbrace =\lbrace 0.17,0.11 \rbrace$ and $\lbrace\langle e_{\geq2} \rangle,\tilde{e}_{\geq2}\rbrace=\lbrace 0.04,0.04\rbrace$, which are clearly smaller than those obtained from the \texttt{6g.60M.5AU} simulations without eccentricity damping: $\lbrace\langle e_{1} \rangle,\tilde{e}_{1}\rbrace =\lbrace 0.25,0.20 \rbrace$ and  $\lbrace\langle e_{\geq2} \rangle,\tilde{e}_{\geq2}\rbrace=\lbrace 0.07,0.06\rbrace$. Hence, these values are significantly influenced by tides, as expected, but also maintain a significant dichotomy between single- and multi-transiting systems. This dichotomy signal is stronger than that which arises from the \textit{in-situ} self scattering model \citep[e.g. ][]{2020MNRAS.491.5595P}.

Compared to the Kepler systems examined by \citet{2019AJ....157..198M}, for which $\{\langle e_{1} \rangle,\langle e_{\geq 2} \rangle\}=\{0.21,0.05\}$, we see that the simulations with tides applied now produce moderately smaller values of the mean eccentricities compared to the Kepler systems. We note, however, that \citet{2019AJ....157..198M} list the seven most eccentric single-transiting systems in their table~1, and five out the seven have orbital periods in excess of 16 days. Furthermore, six out of the seven planets have radii $\ge 2.2$~R$_{\oplus}$, such that the appropriate value of $Q_{\mathrm{p}}$ may significantly exceed our adopted value of $Q_{\mathrm{p}}=100$ if these planets have significant gas envelopes. Hence, it seems likely that the most eccentric Kepler systems identified by \citet{2019AJ....157..198M} have orbital and physical parameters that render tidal eccentricity damping relatively ineffective over Gyr timescales. \citet{2019AJ....157..198M} consider the possibility that the single-transiting systems are actually composed of two sub-populations: low and high eccentricity systems. Our results support this hypothesis, and show that these two populations can be explained by tidal eccentricity damping.

\section{Discussion and conclusions}\label{sec:discuss}
We have presented the results of $N$-body simulations of outer systems of giant planets that coexist with inner compact systems of super-Earths. The outer planetary systems are set up to become dynamically unstable, and the purpose of this study is to examine whether or not perturbations from the outer planets, acting on the inner systems, are able to generate dichotomies in the multiplicity and eccentricity distributions that agree with the Kepler data for compact systems of super-Earths.

There have been previous studies of the influence of outer giant planets perturbing inner systems \citep{2017MNRAS.468.3000M, 2017AJ....153..210H}, and in this work we have examined the influence of the multiplicities, orbital radii and masses of the planets that make up the outer systems. We have considered systems of $N_{\rm out}=3$, 6 and 12 outer planets, that are centred on orbital radii $\tilde{a}_{\rm out}=2$, 5 and 10~au. The planet masses were varied between $M_{\rm p,out}=15$ and 100 Earth masses, and we chose parameters that ensure these bodies would be below the detection thresholds of long-term RV surveys. The inner systems were set up using eight of the known 5-planet systems that were discovered by Kepler as templates. We ran 100 simulations for each combination of inner and outer system templates considered.

The final multiplicities of the inner systems are found to be highly dependent on the architectures of the outer systems. The multiplicity tends to be smaller when the initial values of $N_{\rm out}$ and $M_{\rm p,out}$ increase and $\tilde{a}_{\rm out}$ decreases. Some outer system templates fail to induce significant perturbations on the inner systems, leaving the multiplicity unchanged (`unperturbed systems'), while other outer system templates induce very strong perturbations, leading to instabilities in the inner systems in all runs performed for that template. 

For outer systems with $N_{\rm out}=3$ and $\tilde{a}_{\rm out}=10~{\rm au}$, the gravitational scatterings among the outer planets are not strong enough for any of the giant planets to come close enough to the inner systems to generate noticeable disturbances. On the other hand, the $N_{\rm out}=12$ systems generated strong perturbations on the inner systems, and only the subset of runs with $\tilde{a}_{\rm out}=10~{\rm au}$ allows some systems to survive relatively unperturbed. Except for the \texttt{12g.15M.10AU} template, all other $N_{\rm out}=12$ templates have an $\sim 80~{\rm per~cent}$ chance of completely destroying the inner system, while the other $\sim 20~{\rm per~cent}$ mostly have one inner planet surviving. Hence, the occurrence rates of inner systems of planets that arise from the $N_{\rm out}=12$ runs are much smaller than implied by the Kepler data for planets around Solar-type stars \citep{2013ApJ...766...81F, 2013PNAS..11019273P}. A recent analysis by \citet{2018ApJ...860..101Z} suggests the mean multiplicity of super-Earth systems with periods $<100$ days is $\sim 3$, and approximately $1/3$ of Sun-like stars host compact planetary systems. The $N_{\rm out}=12$ outer systems we have considered produce outcomes that are in clear disagreement with these numbers.

We selected eight outer system templates with $N_{\rm out}=6$ at $\tilde{a}_{\rm out}=5$ or 10~au for a more in-depth investigation of their effects on the inner systems. The results show a wide range of final multiplicities, from the template \texttt{6g.15M.10AU} resulting in unperturbed inner systems in $\sim75$ \% of the runs, to \texttt{6g.100M.5AU} completely destroying $\sim75$ \% of the inner systems. A common outcome is the generation of single planet inner systems with relatively high eccentricities, as required by the observed eccentricity dichotomy. The simulations also produce a clear relation between the final multiplicities and the eccentricity distributions, where inner systems with higher final multiplicities have lower eccentricities. For runs where the final planet number in the inner system $N_{\rm in,final}=5$, the final eccentricity distribution is similar to the initial distribution, as expected. The $N_{\rm in,final}=1$ systems, however, have large eccentricities with a mean of $\langle e_{1} \rangle =0.48$. 

We undertook synthetic transit observations of the final inner systems. We counted the relative numbers of one-planet, two-planet, ..., five-planet systems detected by the synthetic observations, and compared this to the ratios of the observed multiplicities in the Kepler data. We found that the simulated systems in this paper produce better agreement with the Kepler data compared to the multiplicity ratios obtained from $N$-body simulations that only consider self-excitation of inner systems of super-Earths \citep[e.g.][]{2020MNRAS.491.5595P}. Some of our outer templates, such as those with $\tilde{a}_{\rm out}=5~{\rm au}$, resulted in inner systems that are in very good agreement with the Kepler multiplicity ratios, and reproduce the Kepler multiplicity dichotomy.

Synthetic observation of the simulated inner systems produces a very clear eccentricity dichotomy. The single transiting planet systems always have a significantly higher mean eccentricity, $\langle e \rangle$, than the multi-transit systems. Compared to control runs which were performed for inner systems without any outer giant planets, the differences between the mean eccentricities for single systems, $\langle e_{1} \rangle$, and those for multiple systems, $\langle e_{\geq 2} \rangle$, are much larger in our models with outer planets. The control runs produce $\{\langle e_{\geq 2} \rangle, \langle e_{1} \rangle\}=\{0.05, 0.08\}$, whereas we obtain $0.06 \le \langle e_{\geq 2} \rangle \le 0.1$ and $0.11 \le \langle e_{1} \rangle \le 0.25$ for the runs with outer systems. The outer system comprised of 6 planets centred around semi-major axis 5~au with masses of 60 M$_{\oplus}$ resulted in an eccentricity dichotomy characterised by $\{\langle e_{\geq 2} \rangle, \langle e_{1} \rangle\}=\{0.07, 0.25\}$, which is in decent agreement with the values $\{\langle e_{\geq 2} \rangle, \langle e_{1} \rangle\}=\{0.04, 0.3\}$ reported by \citet{2016PNAS..11311431X} and the values
$\{\langle e_{\geq 2} \rangle, \langle e_{1} \rangle\}=\{0.05, 0.21\}$ reported by \citet{2019AJ....157..198M}. While this is the outer system template that gives the best agreement with the Kepler data, other templates resulted in similar eccentricity dichotomies One feature of our study, however, is that the mean eccentricities for multiple transiting systems coming out of the simulations are always larger than reported for the Kepler data. This may in part be due to our choice of initial conditions for the inner system templates, but it is also influenced by the fact that systems of 2 transiting planets contribute significantly to the overall eccentricity distributions of the multiple transiting systems. The synthetically observed 2 transiting planet systems often come from underlying systems that have been significantly perturbed by the outer systems, such that the mean eccentricity increases above $\langle e_{\geq 2} \rangle=0.05$.

We ran some additional simulations to test the effect of including additional physics in the models. Simulations that adopted a realistic collision model, instead of the simple hit-and-stick model adopted in the main suite of simulations, produced results that are very similar to the original simulation set. Hence, we conclude that the collision model is not important for determining the outcome of the simulations. We also considered the effect of eccentricity damping due to tidal interaction with the central star over Gyr timescales, and here we observe a significant change in the eccentricity distributions of both the single- and multi-transiting systems. In spite of this change, we still maintain a significant dichotomy between the final eccentricity distributions for the single- and multi-transiting systems, such that the mean eccentricities obtained are $\lbrace\langle e_{\geq2} \rangle,\langle e_{1} \rangle\rbrace=\lbrace 0.04,0.17\rbrace$, similar to but slightly smaller than the values $\lbrace\langle e_{\geq2} \rangle,\langle e_{1} \rangle\rbrace=\lbrace 0.05,0.21\rbrace$ for the Kepler systems obtained by \citet{2019AJ....157..198M}.  

Finally, we remark that although we have undertaken a wide ranging parameter study of outer planetary systems influencing inner systems of super-Earths, what we have presented here is far from exhaustive in terms of multiplicity, planet mass and semi-major axes for the outer systems. For example, we have considered outer planetary systems consisting of identical planets in terms of mass and radius, and this might influence the outcome in terms of collisions versus ejections \citep{2020MNRAS.491.1369A}. Furthermore, we have also assumed that the inner systems are fully formed at the time when the outer giant planets undergo dynamical instability, and it is possible that in a number of systems the instability occurs earlier during the epoch of formation. These considerations may lead to different outcomes, such that the multiplicity and eccentricity dichotomies can be more accurately reproduced by models than has been achieved in this work. Nonetheless, the study presented here shows that the general scenario of outer unseen planets perturbing inner planetary systems is a promising mechanism for explaining some aspects of the Kepler data.

\section*{Acknowledgements}
The authors wish to thank Alessandro Morbidelli and Seth Jacobson for providing their version of the \textsc{symba} $N$-body code, and an anonymous referee whose comments allowed us to significantly improve this paper. Richard Nelson acknowledges support from STFC through the Consolidated Grants ST/M001202/1 and ST/P000592/1. This research utilised Queen Mary's Apocrita HPC facility, supported by QMUL Research-IT\footnote{http://doi.org/10.5281/zenodo.438045}. 



\bibliographystyle{mnras}
\bibliography{Sanson} 



\appendix
\section{Initial eccentricities of the inner systems} \label{app:A1}
\begin{figure}
\centering
\includegraphics[width=1.0\columnwidth]{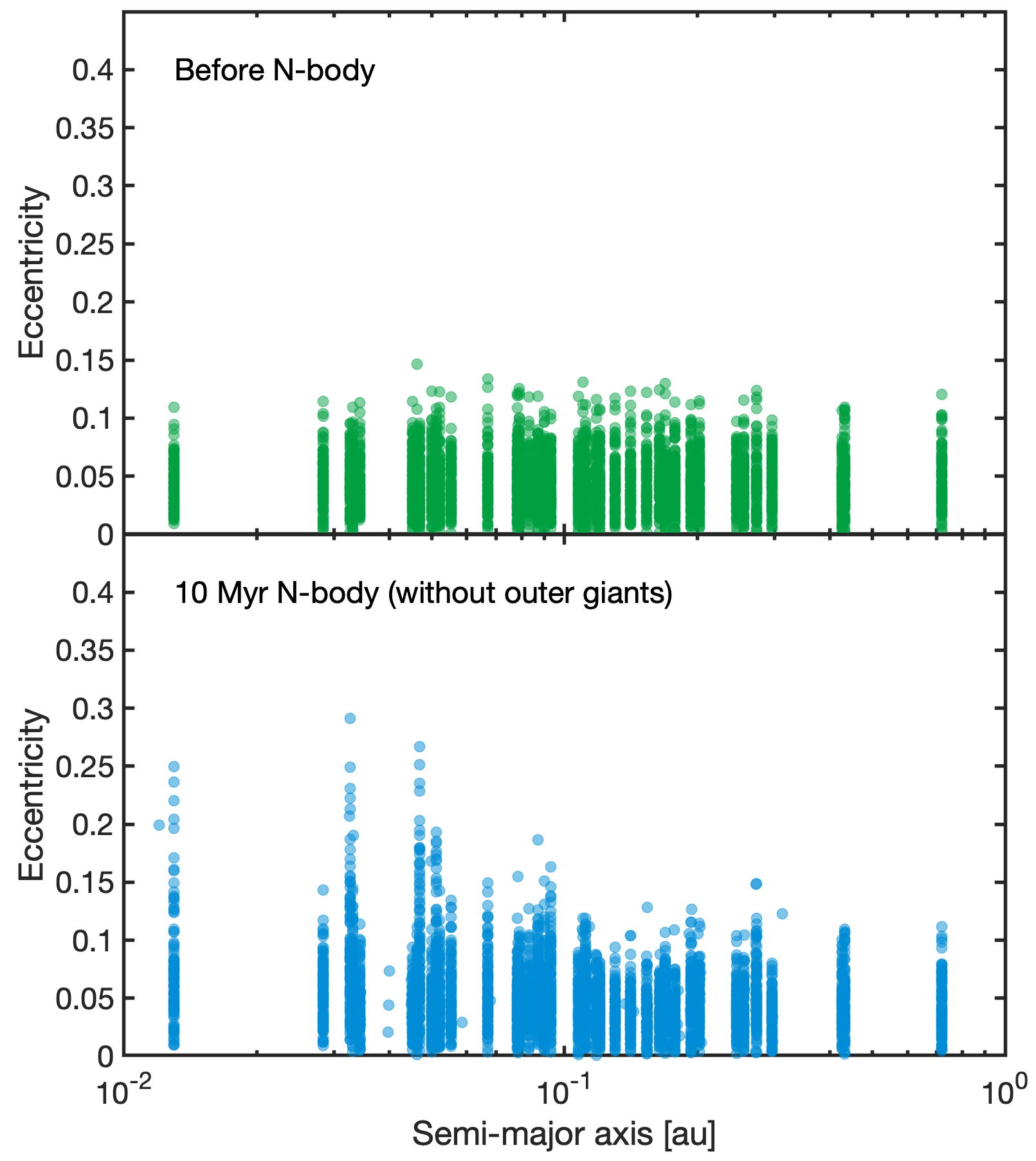}
\caption[]{Eccentricities versus semi-major axis for the inner system before and after they have been evolved for 10 Myr in the absence of giant planets}\label{fig:e_inner}
\end{figure}

\begin{figure}
\centering
\includegraphics[width=1.0\columnwidth]{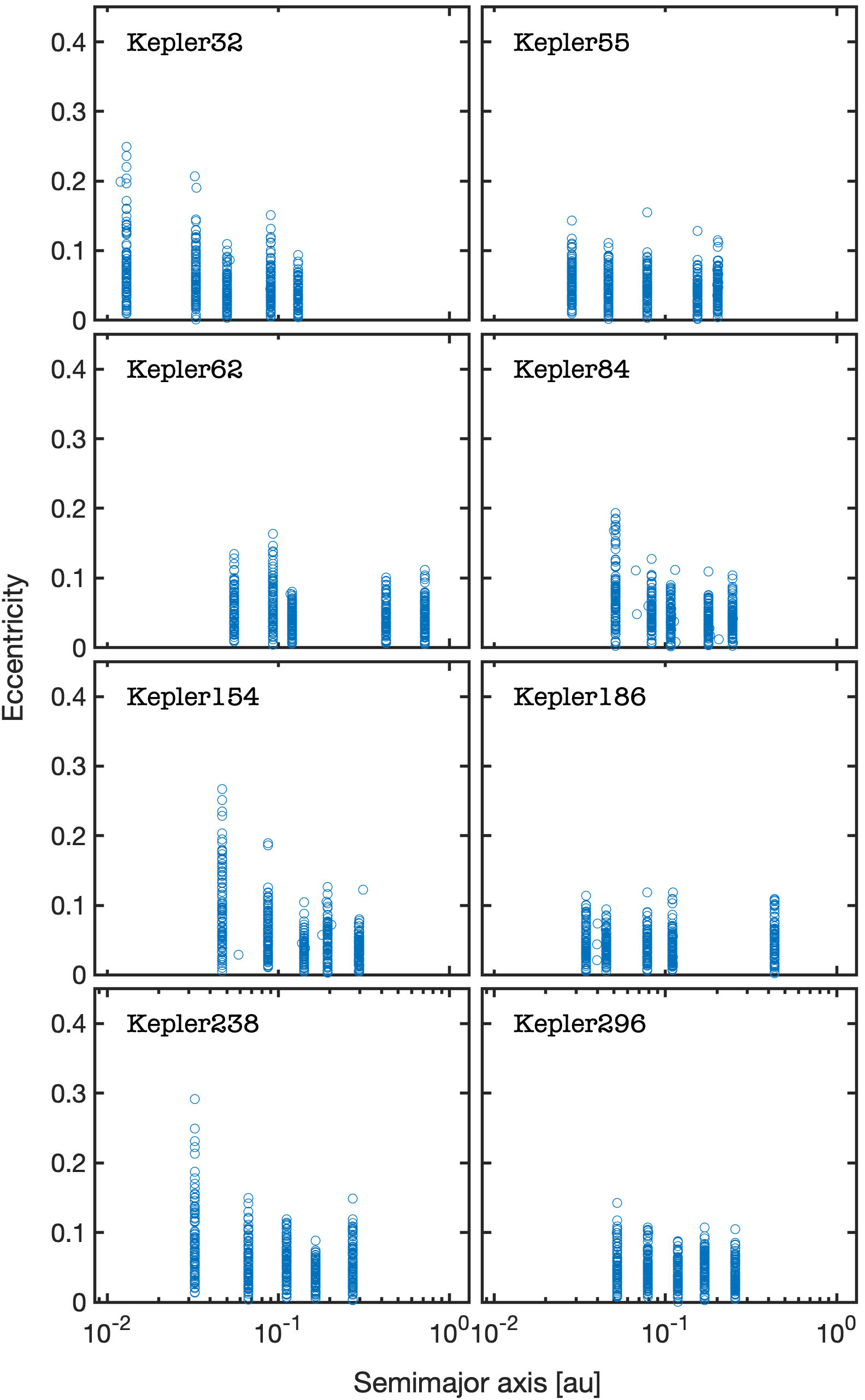}
\caption[]{Eccentricities versus semi-major axis for the inner systems after they have been evolved for 10 Myr in the absence of giant planets}\label{fig:e_inner_per_system}
\end{figure}

The upper panel of figure~\ref{fig:e_inner} shows the initial eccentricities for all planets contained in the inner systems considered in this study. These are the values that arise when setting the eccentricities according to a Rayleigh distribution, as described in Section \ref{subsec:initialcond}. 
The lower panel of figure~\ref{fig:e_inner} shows the final eccentricities after the inner planet systems have been evolved for 10 Myr in the absence of the outer systems of giant planets, and we see that some planets have experienced eccentricity growth even though scattering between the planets has been minimal. 
Figure~\ref{fig:e_inner_per_system} shows the same final eccentricities after the inner planet systems have been evolved for 10 Myr in the absence of the outer systems of giant planets, but now on a system by system basis. In some systems there has been a redistribution of the angular momentum deficit (AMD) causing some of the innermost and lowest mass planets to experience increases in their eccentricities. 

\section{Eccentricity distributions of inner systems from 8 outer system templates obtained by synthetic observations}\label{app:ap}
\begin{figure*}
\centering
\includegraphics[width=0.70\textwidth]{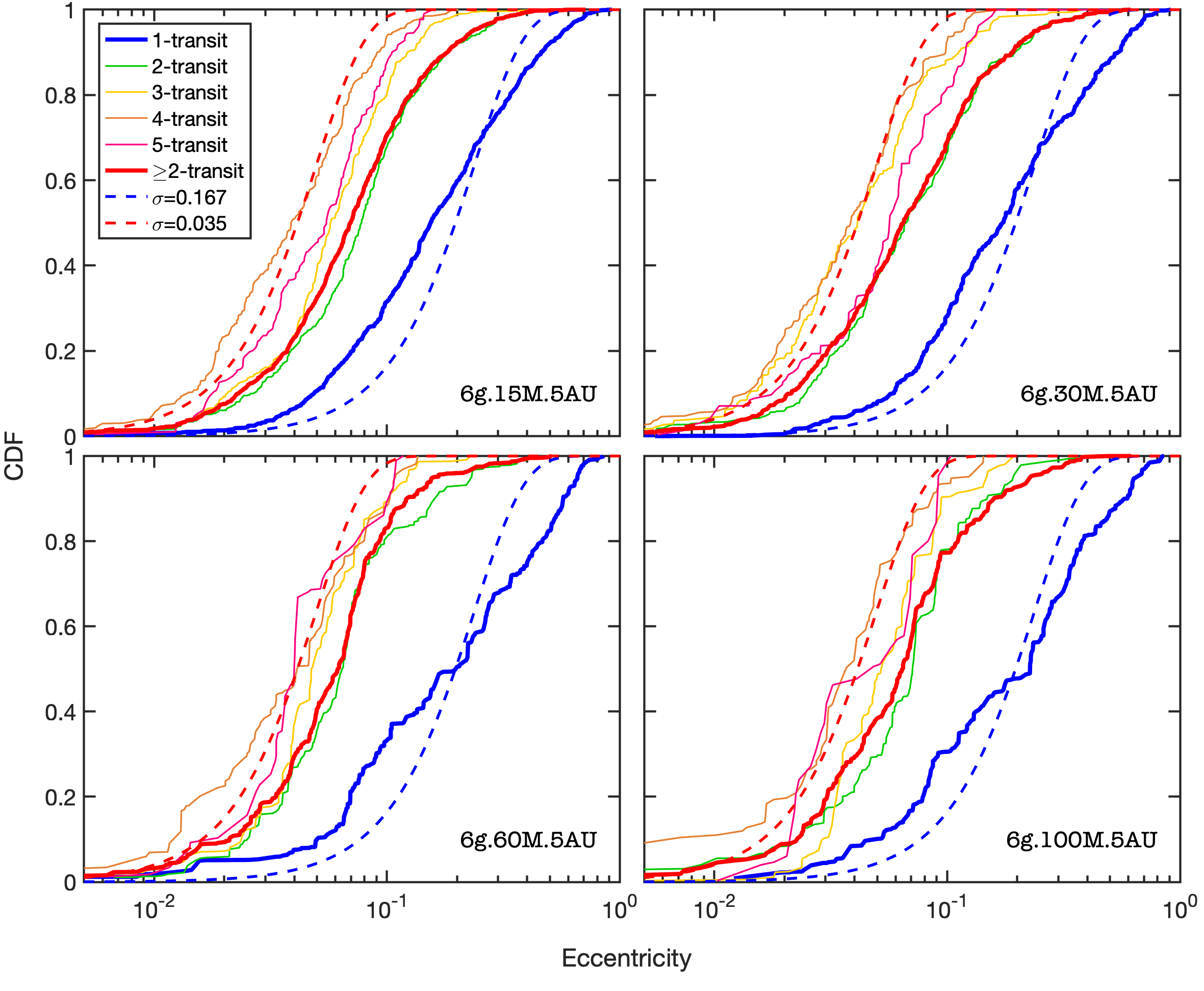}
\caption[Synthetic observed eccentricities CDFs]{CDFs of the eccentricities obtained from the synthetic observation of all outer system templates with $a_{\rm out}=5$ au. For comparison, the CDFs of eccentricities drawn from Rayleigh distributions with eccentricity parameters $\sigma=0.167$ and $\sigma=0.035$ are also plotted.}\label{fig:CDF5AU}
\end{figure*}

\begin{figure*}
\centering
\includegraphics[width=0.70\textwidth]{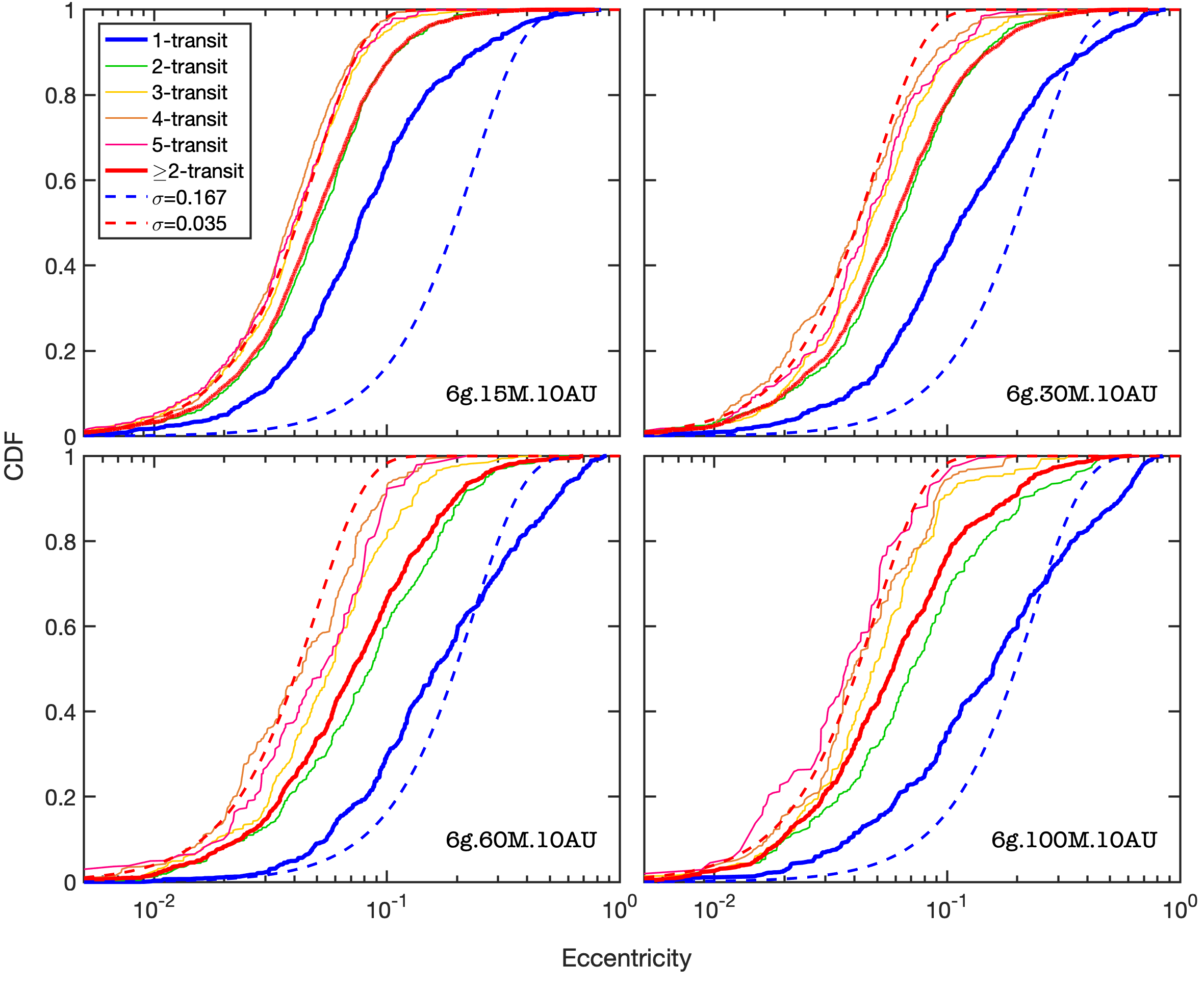}
\caption[Synthetic observed eccentricities CDFs]{CDFs of the eccentricities obtained from the synthetic observation of all outer system templates with $a_{\rm out}=10$ au. For comparison, the CDFs of eccentricities drawn from Rayleigh distributions with eccentricity parameters $\sigma=0.167$ and $\sigma=0.035$ are also plotted.}\label{fig:CDF10AU}
\end{figure*}


\bsp	
\label{lastpage}
\end{document}